\documentclass[aip, jap,
 amsmath,amssymb,
preprint,%
]{revtex4-1}

\usepackage{graphicx}
\usepackage{dcolumn}
\usepackage{bm}
\usepackage{xcolor,soul,framed} 

\usepackage[utf8]{inputenc}
\usepackage[T1]{fontenc}
\usepackage{mathptmx}
\usepackage{etoolbox}

\usepackage{fancyhdr}
\newcommand{\manuscriptfooter}{%
  \fancyhf{}%
  \fancyhead[C]{\footnotesize This article may be downloaded for personal use only. Any other use requires prior permission of the author and AIP Publishing. This article appeared in J. Appl. Phys. 140, 183103 (2026) and may be found at https://doi.org/10.1063/5.0351187.}%
  \fancyfoot[R]{\footnotesize\thepage}%
  \renewcommand{\headrulewidth}{0pt}%
}
\fancypagestyle{manuscript}{\manuscriptfooter}
\fancypagestyle{titlepage}{\manuscriptfooter}
\usepackage{url}

\usepackage{color,soul}
\newcommand\tblu[1]{\textcolor[rgb]{0.00,0.00,0.98}{#1}} 

\renewcommand{\tblu}[1]{{#1}}

\makeatletter
\def\@email#1#2{%
 \endgroup
 \patchcmd{\titleblock@produce}
  {\frontmatter@RRAPformat}
  {\frontmatter@RRAPformat{\produce@RRAP{*#1\href{mailto:#2}{#2}}}\frontmatter@RRAPformat}
  {}{}
}%
\makeatother

\begin{document}

\title[Metasurfaces with Anisotropic Electric and Magnetic Conductivities: Parameter Retrieval for Arbitrary Incidence and Polarization]{Metasurfaces with Anisotropic Electric and Magnetic Conductivities: Parameter Retrieval for Arbitrary Incidence and Polarization}
\author{Dimitrios Chatzidimitriou}
\email{chatzid@eie.gr}
\affiliation{ 
Theoretical and Physical Chemistry Institute, National Hellenic Research Foundation, GR-11635 Athens, Greece
}
\affiliation{
School of Science and Technology, Anatolia American University, 55535, Thessaloniki, Greece.
}
\author{Alexandros Pitilakis}
\affiliation{ 
Theoretical and Physical Chemistry Institute, National Hellenic Research Foundation, GR-11635 Athens, Greece
}
\affiliation{
School of Electrical and Computer Engineering, Aristotle University of Thessaloniki, 54124 Thessaloniki, Greece.
}
\author{Odysseas Tsilipakos}
\email{otsilipakos@eie.gr}
\affiliation{ 
Theoretical and Physical Chemistry Institute, National Hellenic Research Foundation, GR-11635 Athens, Greece
}
\date{\today}

\begin{abstract}
Metasurfaces come in a great variety of unit cell geometries and material composition. A common approach for describing them relies on retrieving effective surface parameters (e.g., electric and magnetic conductivities) from plane-wave scattering data. Conventional retrieval methods are often limited to isotropic unit cells, a uniform host material, or lead to increased complexity by requiring reflection and transmission coefficients from both sides of the metasurface. Here, we derive analytical expressions for the equivalent surface conductivities of an electrically- and magnetically-polarizable anisotropic metasurface under oblique plane-wave incidence with arbitrary polarization. The proposed formulation does not require bidirectional scattering information, thus simplifying the process especially in experimental studies. The method is validated through representative electric and magnetic meta-atoms at microwave and THz frequencies, respectively. Their ability for polarization conversion is demonstrated and discussed relying on the physical insight provided by the retrieved anisotropic conductivities. In addition, we present the inverse expressions for calculating the reflection and transmission coefficients from known surface conductivity tensors, which can prove useful for metasurface synthesis and radiation problems.
\end{abstract}

\maketitle

\section{\label{sec:Intro}Introduction}

Metasurfaces (MSs) have attracted significant attention as a platform for manipulating electromagnetic waves through subwavelength structuring. Research interest started growing exponentially when seminal works demonstrated that despite their subwavelength thickness, these planar arrangements of engineered inclusions can provide abrupt and spatially modulated phase discontinuities enabling wavefront manipulation operations such as beam steering and focusing \cite{Yu2011,Ni2012}. Since then, metasurfaces have been exploited for a wide range of functionalities \cite{Genevet2017,Glybovski2016} ranging from polarization control \cite{Pitilakis2022} and dispersion engineering \cite{Tsilipakos2021,Tsilipakos2020} to light emission, nonlinear phenomena, and nonreciprocal devices \cite{Ramaccia:2020,Nousios2026,Jin2021,Theodosi2025}. 

The response of uniform metasurfaces can be concisely described through homogenized effective parameters (electric and magnetic response functions). This allows to abstract the geometry and material composition of different implementations and focus on the underlying physics; thus, parameter retrieval is customarily used to aid the analysis and design processes. \tblu{The effective parameters can be surface susceptibilities \cite{Holloway2011,Achouri2015,Nitas2023,Cuesta2025}, surface conductivities \cite{Tassin2012,Tsilipakos2018,Orazbay2024}, or meta-atom polarizabilities \cite{Karamanos2014,Asadchy2014}.
In the literature, several works have addressed the problem of retrieving surface susceptibilities or conductivities \cite{Holloway2011,Tassin2012,Achouri2015,Pfeiffer2016, Albooyeh2016, Cuesta2025}.} Indicatively, in Ref.~\onlinecite{Holloway2011} the authors considered diagonal susceptibility tensors (no polarization conversion capabilities). 
\tblu{In Refs.~\onlinecite{Pfeiffer2016, Albooyeh2016, Cuesta2025}, a significantly more general case was studied, involving a fully anisotropic metasurface with magneto-electric coupling (bianisotropy). Specifically in Ref.~\onlinecite{Pfeiffer2016}, although the analysis is comprehensive,} the resulting retrieval formulas are rather complex and require knowledge of reflection and transmission coefficients from both sides of the metasurface. 
The need for bidirectional scattering information and the algebraic complexity of the resulting expressions could potentially limit their practical applicability. 
\tblu{Retrieval from the S-parameters can be directly applied to experimental studies, but other approaches are also useful, including field-averaging techniques \cite{Silveirinha2007} which can be applied to the eigenmode\cite{Nitas2023}.} 

In this work, we derive analytical expressions for retrieving the equivalent electric and magnetic surface conductivities of general anisotropic metasurfaces (nonzero off-diagonal terms) from plane-wave scattering coefficients. 
The presence of artificial magnetic polarizability stems from conduction/displacement current loops in the finite thickness of the MS; \tblu{the only limitations are that the MS is sufficiently thin such that no normal current components develop in its volume and that magnetoelectric coupling is negligible: Magneto-electric coupling becomes possible when the mirror symmetry about the metasurface midplane breaks and is more prominent for electrically-thicker metasurface realizations.} The proposed formulation is elegant and leads to analytic retrieval expressions. Compared to existing approaches, the derived formulas are simpler and require scattering data only from one side of the metasurface, which is advantageous for both practical measurements and theoretical analysis. 
In addition, we present the inverse expressions for calculating the reflection and transmission coefficients from known surface conductivity tensors; they can prove useful for metasurface synthesis and for radiation problems.

In order to showcase the capabilities of the proposed formulation, the retrieval equations are applied to two examples of anisotropic metasurfaces for efficient polarization conversion. We focus on the microwave and THz regions and utilize archetypical electric and magnetic meta-atoms, i.e., a metallic cut-wire and a cut-wire pair, respectively. They are tilted within the unit cell (azimuthally rotated in the lateral plane) in order to produce cross-polarized scattering.
The polarization conversion capabilities in the studied examples is discussed by relying on the physical insight provided by the retrieved conductivity tensors, thus highlighting the merits of the proposed formulation as a metasurface analysis tool. 

The paper is organized as follows: Section~\ref{sec:Deriv} presents the theoretical framework and derives the surface conductivity retrieval formulas. Section~\ref{sec:Results} applies the theory to examples of electric and magnetic meta-atoms and discusses the insights obtained from the retrieved conductivities. The complete derivation as well as additional results and discussions can be found in the Appendix. Section~\ref{sec:Conclusions} concludes the paper.

\section{\label{sec:Deriv}Derivation of retrieval expressions for surface conductivities}

\begin{figure}[!t]
  \begin{center}
  \includegraphics[width=2.5in]{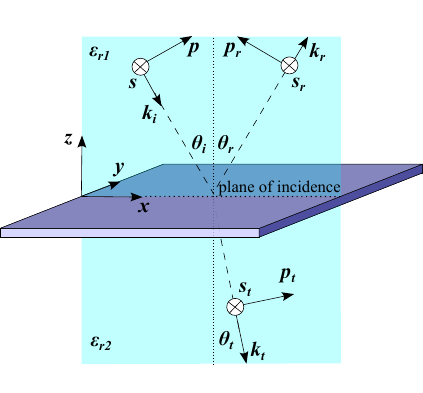}\\
  \caption{Geometry of the surface conductivity retrieval problem. Unit vectors $\mathbf{p}$ and $\mathbf{s}$ denote $p$ (transverse magnetic, TM) and $s$ (transverse electric, TE) polarizations, respectively. Subscripts $i$, $r$, and $t$ denote ``incident'', ``reflected'', and ``transmitted'', respectively. $\mathbf{k}_i, \mathbf{k}_r, \mathbf{k}_t$ are the wavevectors of the incident, reflected and transmitted plane waves, respectively. The incident angle $\theta_i$ and transmitted angle $\theta_t$ are connected through Snell's law. The ultrathin metasurface is located at $z=0$ and the semi-infinite input and output spaces are characterized by permittivity $\varepsilon_1$ and $\varepsilon_2$, respectively.}\label{fig:fig1}
  \end{center}
\end{figure}

A linearly polarized plane wave, propagating in a semi-infinite space with permittivity $\varepsilon_1$, is incident at an oblique angle $\theta_i$ on a two-dimensional (infinitesimal thickness) metasurface located in the $z=0$ plane, Fig.~\ref{fig:fig1}. \tblu{The incoming wave with amplitude $\mathbf{E}_i$ is reflected in the specular direction and transmitted in  medium 2 (permittivity $\varepsilon_2$) at an angle $\theta_t$, given by Snell’s law.Thus, the following discussion is valid for the zeroth (specular) diffraction order.}
The reflected and transmitted fields are also plane waves, which can be written in the TE ($s$) and TM ($p$) component basis as
\begin{subequations}\label{eq:fields}
    \begin{align}                       \mathbf{E}_r&=\overline{\overline{r}}\mathbf{E}_{i}=\left(r_{ss}E_s+r_{sp}E_p\right)\mathbf{\hat{s}}+\left(r_{ps}E_s+r_{pp}E_p\right){\mathbf{\hat{p}}}_r,\\
    \mathbf{E}_t&=\overline{\overline{t}}\mathbf{E}_{i}=\left(t_{ss}E_s+t_{sp}E_p\right)\mathbf{\hat{s}}+\left(t_{ps}E_s+t_{pp}E_p\right){\mathbf{\hat{p}}}_t,
    \end{align}
\end{subequations}
where $\overline{\overline{r}}$ and $\overline{\overline{t}}$ are the $2\times 2$ reflection and transmission tensors (dyadics), which can have all their four elements nonzero. 
The magnetic field is then found through 
\begin{subequations}\label{eq:m_fields}
\begin{align}
    \label{eq:m_fields_i}
    \mathbf{H}_i=\frac{1}{\omega_0 \mu_0}\mathbf{k}_i\times\mathbf{E}_i,\\
    \label{eq:m_fields_r}
    \mathbf{H}_r=\frac{1}{\omega_0 \mu_0}\mathbf{k}_r\times\mathbf{E}_r,\\
    \label{eq:m_fields_t}
    \mathbf{H}_t=\frac{1}{\omega_0 \mu_0}\mathbf{k}_t\times\mathbf{E}_t,
\end{align}
\end{subequations}
where $\mathbf{k}_i$, $\mathbf{k}_r$, and $\mathbf{k}_t$ are the wavevectors for the incoming, reflected, and transmitted waves, respectively. Detailed expressions for all quantities can be found in the Appendix. The wavenumbers in the upper and lower semi-infinite media are $k_1={\omega}{\sqrt{\varepsilon_1\mu_0}}$ and $k_2={\omega}{\sqrt{\varepsilon_2\mu_0}}$.

Assuming negligible normal current components on the metasurface due to the ultrathin thickness, the boundary conditions can be written as \cite{Idemen1990}
\begin{subequations}\label{eq:BC}
 \begin{align}
        \label{eq:BC_1}
        \mathbf{\hat{z}}\times\left(\mathbf{E}_1-\mathbf{E}_2\right)&=-\dfrac{1}{2}\overline{\overline{\sigma}}_m\left(\mathbf{H}_1+\mathbf{H}_2\right)_\parallel,\\
        \label{eq:BC_2}
        \mathbf{\hat{z}}\times\left(\mathbf{H}_1-\mathbf{H}_2\right)&=+\dfrac{1}{2}\overline{\overline{\sigma}}_e\left(\mathbf{E}_1+\mathbf{E}_2\right)_\parallel,
    \end{align}
\end{subequations}
where $\overline{\overline{\sigma}}_{e,m}$ are the tensorial in-plane electric and magnetic conductivities measured in S and $\Omega$, respectively, and ``$||$'' denotes the components parallel (tangential) to the metasurface.  The total electric fields on either side of the metasurface are $\mathbf{E}_1=\mathbf{E}_i+\mathbf{E}_r$ and $\mathbf{E}_2=\mathbf{E}_t$. Equations~\eqref{eq:BC} state that the electric and magnetic fields experience jumps due to the presence of tangential electric and magnetic surface currents, which are connected with the local (averaged) fields through $\mathbf{J}_{m}=(1/2)\overline{\overline{\sigma}}_m(\mathbf{H}_1+\mathbf{H_2})_\parallel$, $\mathbf{J}_{e}=(1/2)\overline{\overline{\sigma}}_e(\mathbf{E}_1+\mathbf{E_2})_\parallel$, respectively.

By substituting Eqs.~\eqref{eq:fields} into Eqs.~\eqref{eq:BC} and using the definitions of unit vectors, wavevevectors and angles shown in Fig.~\ref{fig:fig1}, we can solve for the electric and magnetic surface conductivities. The detailed derivation procedure can be found in the Appendix (Section~A1). Here, we present the final expressions in Eq.~\eqref{eq:sigma_e} and \eqref{eq:sigma_m}; they can be used for retrieving the effective (homogenized) surface conductivities (constitutive parameters) of any anisotropic metasurface on a substrate for arbitrary incidence angle and polarization state, with the only limitations that magnetoelectric coupling is negligible and no normal current components develop in their volume. We note that Eq.~\eqref{eq:sigma_e} and \eqref{eq:sigma_m} give the Cartesian components of the surface conductivities (i.e., $\sigma_{e}^{xx}, \sigma_{e}^{xy}$ etc.), whereas the fields and wavevectors are defined on the TE ($s$) and TM ($p$) polarization basis. $\zeta_{1}^{s}=\eta_1/\cos{\theta_i}$, $\zeta_{2}^{s}=\eta_2/\cos{\theta_t}$, $\zeta_{1}^{p}=\eta_1\cos{\theta_i}$, and $\zeta_{2}^{p}=\eta_2\cos{\theta_t}$ are the wave impedances in medium 1 and 2, respectively, where {$\eta_{1,2}=\sqrt{\mu_0/\varepsilon_{1,2}}$ is the characteristic impedance of each medium}. \tblu{Finally, note the polarization conversion due to anisotropy: {cross-polarized} reflection and transmission coefficients with index ``$sp$'' are found in the off-diagonal terms of Eq.~\eqref{eq:sigma_e} and Eq.~\eqref{eq:sigma_m}. This highlights the connection between the off-diagonal conductivity terms and cross-polarized reflection and transmission (i.e., polarization conversion). Zero cross terms would lead to zero off-diagonal elements.}

Given their generality, Eqs.~\eqref{eq:sigma_e} and \eqref{eq:sigma_m} are relatively simple: They are cast in a succinct matrix multiplication form and involve reflection and transmission coefficients (scattering parameters) that can be specified by numerical simulations or experimental measurements for any metasurface under study. Also, they allow studying practical scenarios where a substrate is present. Another distinct advantage of the derived formulas is that they require scattering data only from one side of the metasurface, since reciprocity relations inherently link forward and backward propagation. 
This substantially simplifies parameter retrieval in both experiments and simulations. Otherwise, ``backward'' reflection and transmission coefficients would {also need to be involved, as in Ref.~\onlinecite{Pfeiffer2016}.} 

\begin{widetext}
\begin{equation}\label{eq:sigma_e}
\overline{\overline{\sigma}}_e = 2 
        \begin{bmatrix}
        \dfrac{1+r_{pp}}{\zeta_{1}^{p}}-\dfrac{1}{\zeta_{2}^{p}}\dfrac{\cos{\theta_t}}{\cos{\theta_i}}t_{pp} & \dfrac{\cos{\theta_i}}{\zeta_{1}^{p}}r_{ps}-\dfrac{\cos{\theta_t}}{\zeta_{2}^{p}}t_{ps} \\ \\
        -\dfrac{1}{\zeta_{1}^{s}}\dfrac{r_{sp}}{\cos{\theta_i}}-\dfrac{1}{\zeta_{2}^{s}}\dfrac{t_{sp}}{\cos{\theta_i}} & \dfrac{1-r_{ss}}{\zeta_{1}^{s}}-\dfrac{t_{ss}}{\zeta_{2}^{s}}
    \end{bmatrix}
        \begin{bmatrix}
        1-r_{pp}+\dfrac{\cos{\theta_t}}{{\cos{\theta}}_i}t_{pp} & -\cos{\theta_i}r_{ps}+\cos{\theta_t}t_{ps} \\ \\
        \dfrac{r_{sp}+t_{sp}}{\cos{\theta_i}} & 1+r_{ss}+t_{ss}
    \end{bmatrix}^{-1},
\end{equation}

\begin{equation}\label{eq:sigma_m}
\overline{\overline{\sigma}}_m
    =
    2\begin{bmatrix}
        \dfrac{t_{sp}-r_{sp}}{\cos{\theta_i}} & -1-r_{ss}+t_{ss}\\
        1-r_{pp}-\dfrac{\cos{\theta_t}}{\cos{\theta_i}}t_{pp} & -\cos{\theta_i}r_{ps}-\cos{\theta_t}t_{ps}
    \end{bmatrix}
    \begin{bmatrix}
        \dfrac{1}{\zeta_{1}^{s}}\dfrac{r_{sp}}{\cos{\theta_i}}-\dfrac{1}{\zeta_{2}^{s}}\dfrac{t_{sp}}{\cos{\theta_i}} & -\dfrac{1-r_{ss}}{\zeta_{1}^{s}}-\dfrac{1}{\zeta_{2}^{s}}t_{ss} \\ \\
        \dfrac{1+r_{pp}}{\zeta_{1}^{p}}+\dfrac{1}{\zeta_{2}^{p}}\dfrac{\cos{\theta_t}}{\cos{\theta_i}}t_{pp} & \dfrac{1}{\zeta_{1}^{p}}\cos{\theta_i}r_{ps}+\dfrac{1}{\zeta_{2}^{p}}\cos{\theta_t}t_{ps}
    \end{bmatrix}^{-1}.
\end{equation}
\end{widetext}

Equations~\eqref{eq:sigma_e} and \eqref{eq:sigma_m} get further simplified when the host medium is uniform ($\varepsilon_2=\varepsilon_1$), see the Appendix [Eqs.~(A17), (A18)].
Moreover, a more compact and symmetric version of Eqs.~\eqref{eq:sigma_e} and \eqref{eq:sigma_m} can be written if we use the Cartesian components of the reflection and transmission tensors, see the Appendix [Eqs.~(A27), (A28)]. 
One can also invert the retrieval expressions and specify the reflection and transmission tensors for given surface conductivities; see the Appendix (Section A2) for the relations and a detailed derivation.

Finally, since many metasurfaces of interest support only electric surface currents (e.g. metallic implementations with a single metallization layer), it is useful to present a simplified version of Eqs.~\eqref{eq:sigma_e} in the absence of magnetic currents. In this case, the transmission and reflection coefficients are not independent [see Appendix, Eqs.~(A23), (A24)], and we can eliminate either one of them. The electric conductivity in terms of the reflection coefficients is then given by 
\begin{subequations}
\label{eq:sigma_e_only_electric}
\begin{align}
    {\sigma}_{e}^{xx}&=\dfrac{-Y_p\left[\left(\dfrac{\Delta_p}{Y_p}-r_{pp}\right)\left(1+r_{ss}\right)+r_{ps}r_{sp}\right]}{D}, \\
    {\sigma}_{e}^{xy}&=\dfrac{2}{\sqrt{\zeta_1^s\zeta_1^p}D}r_{ps}, \\
    {\sigma}_{e}^{yx}&=-\dfrac{2}{\sqrt{\zeta_1^s\zeta_1^p}D}r_{sp}, \\
    {\sigma}_{e}^{yy}&=\dfrac{-Y_s\left[\left(\dfrac{\Delta_s}{Y_s}+r_{ss}\right)\left(1-r_{pp}\right)+r_{ps}r_{sp}\right]}{D},
\end{align}
\end{subequations}
where we have introduced the conductivities $Y_j=1/\zeta_{1}^j+1/\zeta_{2}^j$ and $\Delta_j=1/\zeta_{2}^j-1/\zeta_{1}^j$ with $j=\{s,p\}$, and the factor $D=\left(1+r_{ss}\right)\left(1-r_{pp}\right)+r_{sp}r_{ps}$. For the simplest case of an isotropic metasurface in a uniform host medium, Eqs.~\eqref{eq:sigma_e_only_electric} reduce to 
\begin{subequations}
\label{eq:sigma_e_only_electric_isotrop}
\begin{align}
    {\sigma}_{e}^{xx}&=\dfrac{2}{\zeta^p}\dfrac{r_{pp}}{1-r_{pp}}, \\
    {\sigma}_{e}^{yy}&=-\dfrac{2}{\zeta^s}\dfrac{r_{ss}}{1+r_{ss}}.
\end{align}
\end{subequations}


%
%
%
%
%
%
%
%
%
%
%
%
%

\section{Application to electric and magnetic cells\label{sec:Results}}

In this section, we apply the retrieval equations in two characteristic metallic metasurface examples, one exhibiting purely electric polarizability and another which is magnetically-polarizable as well. We employ numerical simulations (CST Studio) to compute the plane-wave reflection and transmission spectra (S-parameters) and calculate the effective (homogenized) electric and magnetic conductivities through Eqs.~\eqref{eq:sigma_e} and \eqref{eq:sigma_m}. By comparing the conductivity components with the scattering response, we can capture additional physical insight. Note that the self-consistency of the process is validated by feeding the retrieved conductivities back into Eqs.~(A20) and (A22) (inverse expressions). \tblu{The simulation setup is a typical one for periodic structures: Floquet periodic boundary conditions at the side walls and two ports (input and output) below and above the structure backed by perfectly matched layers to suppress unwanted reflections. A typical experimental setup at microwave frequencies, for approximately measuring plane-wave scattering coefficients, would involve horn antennas of appropriate directivity at sufficient (farfield) distance from the sample. For an example in reflection geometry, see for instance Ref.~\onlinecite{Tsilipakos2021}.}

\tblu{We should note that the retrieval expressions do not depend on how the reflection and transmission coefficients are obtained. Here we used the finite element method, a rigorous general-purpose full-wave numerical method, but they could also be supplied by measurements or by other numerical methods, including integral-equation approaches. The latter have been employed, for example, for the design of stacked metasurfaces \cite{Budhu2022} and for inverse scattering problems involving the material properties of an illuminated object \cite{Valagiannopoulos2010}. 
}


\subsection{\label{sec:Electric}Electric meta-atom for THz frequencies}

\begin{figure}[]
  \begin{center}
  \includegraphics[width=85mm]{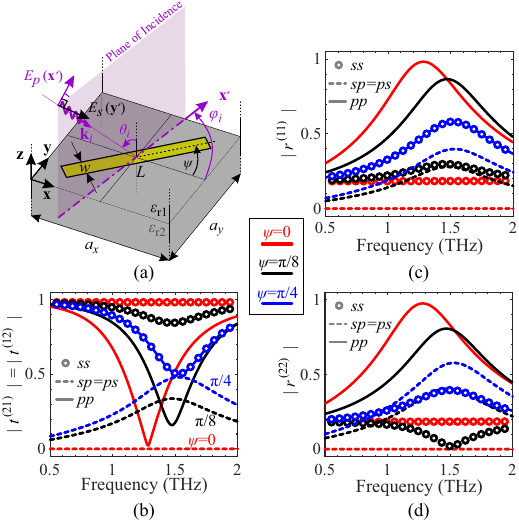}\\
  \caption{(a)~Schematic of the tilted cut-wire resonator lying at the interface of two isotropic dielectrics ($\varepsilon_{r1}=1$ and $\varepsilon_{r2}=2.1$). (b)~Transmission spectra at normal incidence in the $xz$ plane $(\theta_i,\varphi_i)=(0,0)$ for different in-plane rotations, $\psi$, of the wire. (c)-(d)~Reflection spectra for illumination from the two sides of the metasurface. The dependence of co- and cross-polarized responses on $\psi$ highlights the anisotropic nature of the meta-atom. For $\psi=\pi/4$ (blue), $ss$ and $pp$ responses coincide and the $sp$ component becomes most pronounced.}
  \label{fig:Elec1RT}
  \end{center}
\end{figure}

\begin{figure*}[]
  \begin{center}
  \includegraphics[width=150mm]{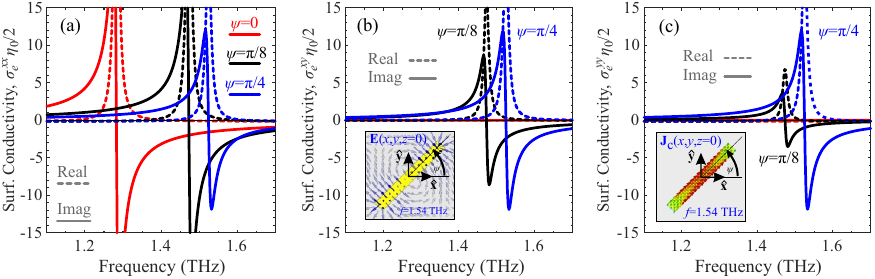}\\
  \caption{Retrieved normalized (dimensionless) electric surface conductivity components of the cut-wire meta-atom at $(\theta_i,\varphi_i)=(0,0)$ illumination for different rotations $\psi$ by utilizing the scattering data in Fig.~\ref{fig:Elec1RT}(b),(d). (a)~$\sigma_e^{xx}$, (b) $\sigma_e^{xy}=\sigma_e^{yx}$, and (c)~$\sigma_e^{yy}$. The conductivity components exhibit Lorentzian-like resonances whose frequency and linewidth are directly associated with the characteristics of the spectral features in the scattering data of Fig.~\ref{fig:Elec1RT}.
  The inset in panel (b) shows the on-resonance electric field distribution across the $z=0$ plane for the case of $\psi=\pi/4$ and the inset in panel (c) shows corresponding conduction current on the cut wire. An electric dipole resonance can be readily identified. }
  \label{fig:Elec2SigmaNormal} 
  \end{center}
\end{figure*}

\begin{figure}[]
  \begin{center}
  \includegraphics[width=85mm]{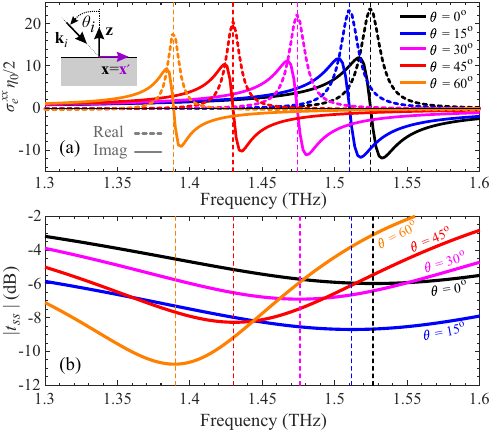}\\
  \caption{(a) Retrieved normalized (dimensionless) electric surface conductivity of the cut-wire meta-atom at $\psi=\pi/4$ for different incidence angles $\theta_i$. Only $\sigma_e^{xx}$ is shown; other components exhibit similar behavior. (b) Corresponding co-polarized transmission spectra. The shift of the resonance with increasing $\theta_i$ is consistently captured by the retrieved conductivity.}
  \label{fig:Elec3SigmaOblique}
  \end{center}
\end{figure}

We first study an archetypical cut-wire meta-atom on a square $a_x=a_y=85~\mu$m lattice, Fig.~\ref{fig:Elec1RT}(a). 
The gold wire with dimensions $w\times L=10~\mu\mathrm{m}\times82~\mu$m (resonating in the 1~THz region) lies on the interface between air and a dielectric substrate ($\varepsilon_r=2.1$). The wire is rectangular, it possesses only $C_2$ rotational symmetry and is inherently anisotropic with respect to the two Cartesian axes. The wire can be rotated on the $xy$ plane around the $z$ axis, as denoted by the azimuth angle $\psi$ [Fig.~\ref{fig:Elec1RT}(a)]. The transmission and reflection spectra (S-parameters) under normal incidence are depicted in Fig.~\ref{fig:Elec1RT}(b)-(d) for three different rotation angles within $\psi\in[0, \pi/4]$. Given that incidence takes place at a principal plane ($\varphi_{i}=0$, or more generally $\varphi_{i}= m\pi/2$ with $m$ integer), cross-polarized scattering is maximized for $\psi=\pi/4$, as would be intuitively anticipated. This can be seen by the $sp$/$ps$ components in Fig.~\ref{fig:Elec1RT}(b)-(d). Fig.~\ref{fig:Elec1RT}(b) depicts transmission coefficients whereas  Fig.~\ref{fig:Elec1RT}(c),(d) depicts reflection coefficients for illumination from both sides of the metasurface. 
On the other hand, when the incident polarization is parallel or perpendicular to the metal wire ($\psi=0$ or $\psi=90^\circ$), {then $|r_{sp}|\rightarrow0$ and $|t_{sp}|\rightarrow0$, meaning that no polarization conversion takes place.} 

Plugging the scattering parameters of Fig.~\ref{fig:Elec1RT}(b),(d) into Eq.~\eqref{eq:sigma_e}, we retrieve the anisotropic electric conductivity tensor, whose components are depicted in  Fig.~\ref{fig:Elec2SigmaNormal}. No magnetic conductivity is expected in this case as the single metallization layer cannot sustain current loops perpendicular to the illuminating magnetic field; this intuition is verified by the retrieval: the $\sigma_m$ spectra computed with Eq.~\eqref{eq:sigma_m} are practically zero (not shown). We stress once more that only the scattering data for a single illumination direction are used, without requiring reflection coefficients from the other side [i.e., the data in Fig.~\ref{fig:Elec1RT}(c)].  The retrieved co-polarized components are depicted in Fig.~\ref{fig:Elec2SigmaNormal}(a),(c). A clear Lorentzian resonance in the electric surface conductivity is observed, as one would expect for the fundamental resonance of a cut-wire which is essentially an electric dipole resonance; this can be also verified by the insets in Fig.~\ref{fig:Elec2SigmaNormal}(b) and Fig.~\ref{fig:Elec2SigmaNormal}(c), which depict electric field and conduction current distributions, respectively, on resonance. For $\psi=0$, the Lorentzian resonance appears near 1.3~THz, consistent with the spectral features in Fig.~\ref{fig:Elec1RT}(b)-(d). As $\psi$ varies, these spectral features shift to higher frequencies, a behavior which is reproduced by the retrieved conductivity. Moreover, the spectral features become weaker, as is also the case with the strength (linewdith) of the resonance. For details on the analytical connection between the coupling strength and damping of a Lorentzian resonance in the surface conductivity and the characteristics of the resulting spectral feature (full-width  at half maximum and peak reflection) the reader is directed to Ref.~\onlinecite{Tsilipakos2021}. 

The  cross-polarized component ($\sigma_e^{xy}=\sigma_e^{yx}$) is depicted in Fig.~\ref{fig:Elec2SigmaNormal}(b). For $\psi=0$, $\sigma_e^{xy}$ is negligible in agreement with $r_{sp}$ and $t_{sp}$ in Fig.~\ref{fig:Elec1RT}. As $\psi$ increases, a Lorentzian resonance in $\sigma_e^{xy}$ appears at the same frequency as the co-polarized components. The strength of the resonance is maximized for  $\psi=\pi/4$, in agreement with the behavior of the $sp$ components in Fig.~\ref{fig:Elec1RT}.

Finally, we study the effect of incidence angle $\theta_i$ (elevation angle within the plane of incidence). The retrieved co-polarized response is plotted in Fig.~\ref{fig:Elec3SigmaOblique}(a). The frequency of the Lorentzian resonance shifts with $\theta_i$ indicating spatial dispersion. This behavior closely follows the trend in the transmission spectra in Fig.~\ref{fig:Elec3SigmaOblique}(b), confirming the consistency of the approach. 

\subsection{\label{Magnetic}Magnetic meta-atom for GHz frequencies}

\begin{figure}[]
  \begin{center}
  \includegraphics[width=85mm]{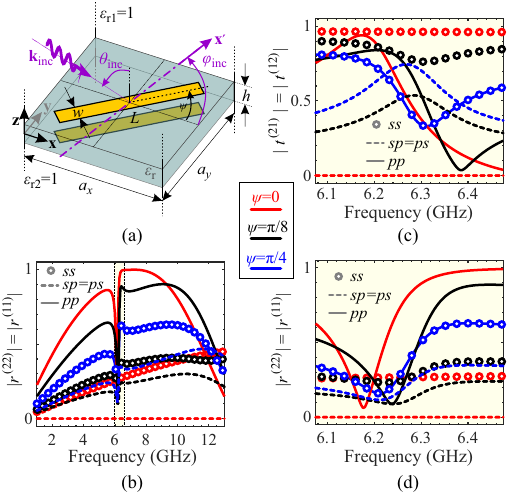}\\
  \caption{(a) Schematic of the tilted cut-wire-pair resonator on a printed circuit board. (b) Wide-band reflection spectra at $(\theta_i,\varphi_i)=(0,0)$ for different in-plane rotations $\psi$. (c)-(d) Narrow-band transmission and reflection spectra focusing in the region of the magnetic resonance.}
  \label{fig:Magn1RT}
  \end{center}
\end{figure}

\begin{figure*}
  \begin{center}
  \includegraphics[width=150mm]{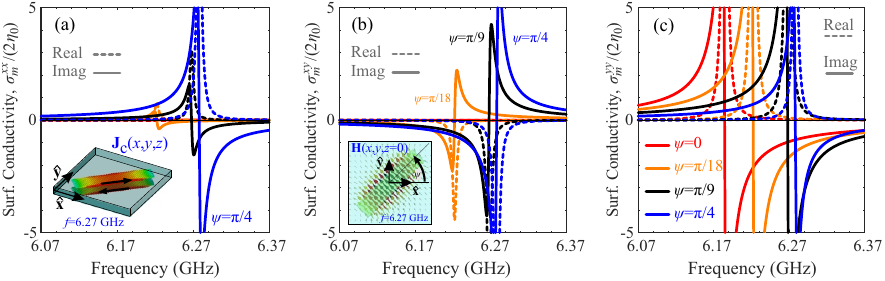}\\
  \caption{Retrieved normalized (dimensionless) magnetic surface conductivity components for the cut-wire-pair meta-atom at $(\theta_i,\varphi_i)=(0,0)$ for different rotations $\psi$. (a)~$\sigma_m^{xx}$, (b)~$\sigma_m^{xy}=\sigma_m^{yx}$, and (c)~$\sigma_m^{yy}$. The retrieved conductivities exhibit Lorentzian-like resonances whose frequency and linewidth are directly associated with the characteristics of the spectral features in the scattering data of Fig.~\ref{fig:Magn1RT}. Insets show the on-resonance distribution of the antiparallel conduction currents [inset in panel (a)] and the magnetic field at the mid-plane $z=0$ [inset in panel (b)] for the case $\psi=\pi/4$. A magnetic dipole resonance can be readily identified.}
  \label{fig:Magn2SigmaNormal}
  \end{center}
\end{figure*}

\begin{figure}[]
  \begin{center}
  \includegraphics[width=85mm]{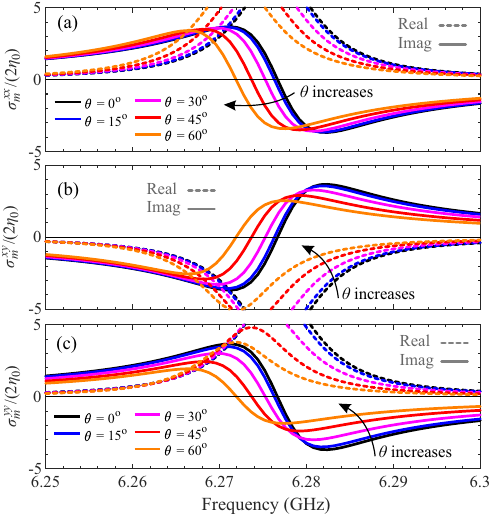}\\
  \caption{Retrieved normalized (dimensionless) magnetic surface conductivity of the cut-wire-pair meta-atom at $\psi=\pi/4$ for different incidence angles $\theta_i$. (a)~$\sigma_m^{xx}$, (b)~$\sigma_m^{xy}=\sigma_m^{yx}$, and (c)~$\sigma_m^{yy}$. Note that $\sigma_m^{xx}$ and $\sigma_m^{yy}$ are equivalent only for normal incidence. The shift in the resonant frequency with incidence angle indicates spatial dispersion.}
  \label{fig:Magn3SigmaOblique}
  \end{center}
\end{figure}

Next, we study a cut-wire-pair meta-atom on a two-metallization-layer printed circuit board, Fig.~\ref{fig:Magn1RT}(a). Cut-wire-pair meta-atoms are archetypical ``magnetic'' meta-atoms in that they can support magnetic resonances by means of antiparallel surface currents on the two wires. The wires are made of copper ($\sigma=5.8\cdot 10^7$~S/m) with dimensions $w\times L=3~\mathrm{mm}\times14~\mathrm{mm}$ patterned on a low-loss board material ($h=1.5$~mm, $\varepsilon_r=3$, $\tan\delta=0.001$; e.g. Rogers RO3003). The host material is uniform ($\varepsilon_{r1}=\varepsilon_{r2}=1$) and the periodicity is square with $a_x=a_y=15$~mm. The structure can be rotated in-plane as denoted by the angle $\psi$.

The reflection spectra for normal incidence are depicted in Fig.~\ref{fig:Magn1RT}(b) for a broad frequency range (1-13 GHz). In contrast to Section~\ref{sec:Electric}, reflection is identical from both sides, since the host medium is uniform.
For the aforementioned parameters, the fundamental magnetic resonance arises in the vicinity of 6~GHz. An electric resonance is also supported (parallel currents along the two stripes), but at higher frequency (8-10~GHz), see Fig.~\ref{fig:SigmeElec_of_CutWirePair}. The sharp magnetic resonance falls within the broad linewidth of the strong electric resonance, thus giving rise to an asymmetric Fano lineshape.
We next focus on the vicinity of the magnetic resonance and depict transmission [Fig.~\ref{fig:Magn1RT}(c)] and reflection [Fig.~\ref{fig:Magn1RT}(d)] coefficients for different rotation angles $\psi$.
For $\psi=0$, the $s$ and $p$ polarizations exhibit distinct responses due to the rectangular {(non-square)} shape of the wire; for $\psi=\pi/4$ they coincide since the $x$ and $y$ axes become equivalent. The off-diagonal scattering parameters (polarization conversion) increase with $\psi$ and are maximized at $\pi/4$, as intuitively anticipated.

Based on these spectra, we retrieve the magnetic surface conductivity tensor using Eq.~\eqref{eq:sigma_m}; the different components are depicted in Fig.~\ref{fig:Magn2SigmaNormal}. As in the previous case, the retrieval is carried out without requiring bidirectional scattering information.
For $\psi=0$, we observe a Lorentzian resonance in $\sigma_m^{yy}$ since the illuminating magnetic field ($H_y$) is parallel to the induced magnetic dipole moment ($m_y$). The magnetic dipole nature of the supported resonance can be verified by the insets in Fig.~\ref{fig:Magn2SigmaNormal}(a) and Fig.~\ref{fig:Magn2SigmaNormal}(b), which depict the antiparallel conduction currents on the two wires and the corresponding magnetic field at the midplane $z=0$. As $\psi$ varies, (i)~the resonance is now imprinted on the $\sigma_m^{xx}$ component as well, since the induced magnetic dipole moment acquires a non-zero projection on the $x$ axis, and (ii)~the off diagonal component ($\sigma_m^{xy}$=$\sigma_m^{yx}$) increases, indicating enhanced polarization conversion. For $\psi=\pi/4$, $\sigma_m^{xx}$ becomes as strong as $\sigma_m^{yy}$, and $\sigma_m^{xy}$ becomes maximum. 
We should note that in Fig.~\ref{fig:Magn2SigmaNormal}(b), $\sigma_m^{xy}$ is antiresonant and the real part is negative. However, this does not violate passivity as would be the case for the diagonal terms of the conductivity tensor, see Section~B in the Appendix. 


Furthermore, we keep $\psi=\pi/4$ fixed (corresponds to the strongest polarization conversion) and apply the retrieval process for a range of incidence angles within the $xz$ plane of incidence ($\varphi_i=0$). Figure~\ref{fig:Magn3SigmaOblique} depicts the elements of the surface conductivity tensor. Note that $\sigma_m^{xx}$ and $\sigma_m^{yy}$ are equivalent only for normal incidence. As $\theta_i$ increases, the  $\sigma_m^{yy}$ component weakens since the projection of the illuminating $H$-field on the $x$ axis decreases.
As with the electric meta-atom, we see a shift of the Lorentzian resonance with $\theta_i$, indicating the presence of spatial dispersion.

\begin{figure}[]
  \begin{center}
  \includegraphics[]{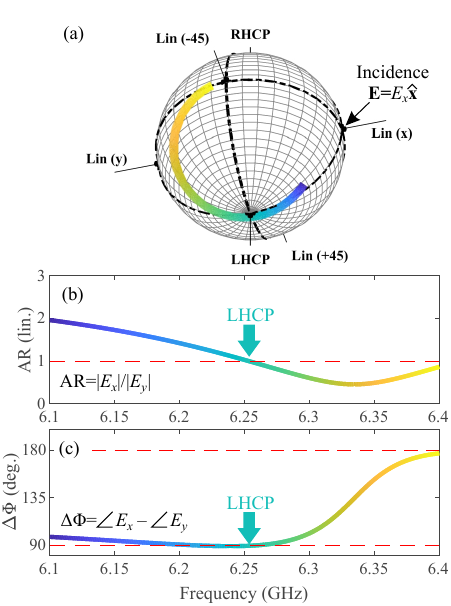}\\
  \caption{{Polarization conversion for a plane wave transmitted through the cut-wire-pair meta-atom (rotation $\psi=\pi/4$) when $\theta=0$ and assuming linear $x$-polarized illumination. (a)~Transmitted polarization state mapped on the Poincar\'{e} sphere for the frequency range 6.1-6.4~GHz. (b)~Axial ratio and (c)~phase difference spectra for the transmitted plane wave. The transmitted polarization is almost perfectly LHCP at 6.25~GHz and it approaches the linear $-45^\circ$ state at 6.4~GHz.}}
  \label{fig:MagnPolRot}
  \end{center}
\end{figure}

Finally, relying on the spectra of Fig.~\ref{fig:Magn1RT}(c) one can observe that co- and cross-polarized transmission coefficients become equal in magnitude for two frequencies near 6.25 and 6.4~GHz ($\psi=\pi/4$ curve). Such regions indicate opportunities for linear-to-circular polarization conversion. We further investigate this in Fig.~\ref{fig:MagnPolRot}, which assesses the output polarization for linear $x$-polarized illumination in the frequency range 6.1-6.4~GHz. Looking at the Poincar\'{e} sphere, the output polarization is almost perfectly LHCP at 6.25~GHz and it approaches the linear $-45^\circ$ state at 6.4~GHz. This can be also seen in Fig.~\ref{fig:MagnPolRot}(b),(c) depicting the axial ratio and phase difference of the two output linear components. Note that at 6.25~GHz a $y$-polarized plane wave will be converted to RHCP, see Fig.~\ref{fig:polarization_conv} for more details.

We should note that the presence of an electric resonance in the vicinity of the magnetic one [see Fig.~\ref{fig:SigmeElec_of_CutWirePair}] has proven essential for achieving the linear-to-circular polarization conversion at 6.25~GHz. More specifically, using our formulation one can arrive at the required surface conductivities for  this to happen. The Appendix contains a detailed derivation of these criteria for two cases: (i)~a purely electric meta-atom (ii)~a purely magnetic meta-atom. Section~D focuses on symmetrically rotated cases ($\psi=\pi/4$), where both orthogonal linear polarizations are converted simultaneously to circular, whereas Section~E focuses on a single linear polarization. For the sake of simplicity in the analytically-derived criteria, we have assumed normal incidence. Here, we highlight the end result for a purely magnetic ($\overline{\overline{\sigma}}_e=0$), symmetrically-rotated ($\sigma_m^{xx}=\sigma_m^{yy}$) meta-atom. From Eq.~(D5b), the necessary condition for converting $s/p$ incident light to LHCP/RHCP is (same dielectric on both sides):
\begin{equation}
\label{eq:magnetic conductivity for polarization conversion_transmission}
    \frac{\sigma_m^{xy}}{2\eta}=\frac{\sigma_m^{yx}}{2\eta}=i(1+\frac{\sigma_m^{xx}}{2\eta}).
\end{equation}

In Fig.~\ref{fig_s:polarization_condition}, we investigate whether the retrieved magnetic conductivity of the tilted wire pair would satisfy this criterion. 
One then finds that Eq.~\eqref{eq:magnetic conductivity for polarization conversion_transmission} is not exactly satisfied for any frequency. This means that linear-to-circular conversion would not have been possible had the meta-atom been purely magnetic. The electric resonance, whose tail is in the vicinity of the magnetic resonance, changes this picture. 
In the Appendix, we also show that the purely electric meta-atom of Section~\ref{sec:Electric} does not achieve linear-to-circular conversion.



\section{\label{sec:Conclusions}Conclusion}

In conclusion, we have presented a parameter retrieval method for determining the effective electric and magnetic surface conductivities of general anisotropic metasurfaces (nonzero off-diagonal elements). The proposed formulation leads to simple analytic expressions for the conductivity tensors and requires scattering data only from a single side of the metasurface. 
In addition, we have presented the inverse expressions for determining the scattered fields from known conductivity tensors, which can find use in metasurface synthesis and radiation problems. 

The method has been validated using two representative meta-atoms (tilted cut-wire and tilted cut-wire pair) exhibiting distinct resonant behavior. In both cases, the retrieved conductivity tensors were shown to reproduce the key features of the scattering response, with Lorentzian resonances providing a clear physical interpretation of the observed spectra. Their ability for polarization conversion has been discussed relying on the physical insight
provided by the retrieved anisotropic conductivities.
Our results establish an analytical framework for facilitating the analysis and design of anisotropic metasurfaces.


\tblu{Finally, note that our formulation is general, but cannot be applied in cases of strong magneto-electric coupling, such as electrically-thick metasurfaces with multiple metallization layers with asymmetric patterns and/or spacer substrates. When in doubt, it is advised that the presence of strong magneto-electric coupling is assessed beforehand. This can be exercised by either far-field scattering (e.g., assessing if there is any optical activity) and/or through the near fields and induced currents (e.g., by calculating the polarizabilities from the induced dipole moments).}


\section*{Author Declarations}
The authors have no conflicts to disclose.

\section*{Data Availability Statement}
The data that support the findings of this study are available from the corresponding author upon reasonable request. 

\begin{acknowledgments}
Support by the European Union under Grant Agreement 101168152 (Project AVALON). Views and opinions expressed are however those of the author(s) only. Neither the European Union nor the granting authority can be held responsible for them.
\end{acknowledgments}



\appendix
\onecolumngrid

\clearpage
\section{Analytical treatment of metasurface with anisotropic electric and magnetic conductivities}

\subsection{Derivation of equations for parameter retrieval under oblique incidence}
In this section, we provide the detailed derivation of the formulae Eq.~(4), (5). The complex amplitude of the incident electric field is given by
\begin{subequations}
    \begin{align}
        \mathbf{E}_{i}&=E_s\mathbf{\hat{s}}+E_p\mathbf{\hat{p}},
    \end{align}
\end{subequations}
and the total field is formed by multiplying with the propagation term $\exp[{i\left(\mathbf{k}_i\cdot\mathbf{r}-\omega t\right)}]$. The reflected and transmitted complex amplitudes will be denoted as $\mathbf{E}_r$ and $\mathbf{E}_t$, respectively, and again the total fields are formed by multiplying with $\exp[{i\left(\mathbf{k}_r\cdot\mathbf{r}-\omega t\right)}]$ and $\exp[{i\left(\mathbf{k}_t\cdot\mathbf{r}-\omega t\right)}]$. The unit vectors $\mathbf{\hat{s}}$ and $\mathbf{\hat{p}}$ correspond to the ``$s$'' (TE) and ``$p$'' (TM) linear polarization of the incident wave (see Fig.~1 for a graphical representation). Assuming that the $xz$ plane is the plane of incidence, the wavevectors are defined as
\begin{subequations}
    \begin{align}
\mathbf{k}_i&=k_1\left(\sin{\theta_i}\mathbf{\hat{x}}-\cos{\theta_i}\mathbf{\hat{z}}\right),\\
\mathbf{k}_r&=k_1\left(\sin{\theta_i}\mathbf{\hat{x}}+\cos{\theta_i}\mathbf{\hat{z}}\right),\\
\mathbf{k}_t&=k_2\left(\sin{\theta_t}\mathbf{\hat{x}}-\cos{\theta_t}\mathbf{\hat{z}}\right),
    \end{align}
\end{subequations}
where $k_1={\omega}{\sqrt{\varepsilon_1\mu_0}}$ and $k_2={\omega}{\sqrt{\varepsilon_2\mu_0}}$ are the wavenumbers in the upper and lower semi-infinite media. The unit vectors defining the field polarization take the form
\begin{subequations}\label{eq_s:unit_vectors}
    \begin{align}
        \mathbf{\hat{s}}&\equiv{\mathbf{\hat{s}}}_r\equiv{\mathbf{\hat{s}}}_t=\mathbf{\hat{y}},\\
        \mathbf{\hat{p}}&={\mathbf{\hat{k}}}_i\times\mathbf{\hat{s}}= \cos{\theta_i}\mathbf{\hat{x}}+\sin{\theta_i\mathbf{\hat{z}}},\\
        {\mathbf{\hat{p}}}_r&={\mathbf{\hat{k}}}_r\times\mathbf{\hat{s}}=-\cos{\theta_i}\mathbf{\hat{x}}+\sin{\theta_i\mathbf{\hat{z}}},\\
        {\mathbf{\hat{p}}}_t&={\mathbf{\hat{k}}}_t\times\mathbf{\hat{s}}= \cos{\theta_t}\mathbf{\hat{x}}+\sin{\theta_t\mathbf{\hat{z}}},
    \end{align}
\end{subequations}
where the unit vectors $\mathbf{\hat{s}}_r$ and $\mathbf{\hat{p}}_r$ refer to the reflected wave, and $\mathbf{\hat{s}}_t$ and $\mathbf{\hat{p}}_t$ refer to the transmitted wave.

Finally, the reflected and transmitted electric and magnetic field amplitudes are given by
\begin{subequations}\label{eq_s:r_t_amplitudes}
    \begin{align}
        \mathbf{E}_r&=\overline{\overline{r}}\mathbf{E}_{i}=\left(r_{ss}E_s+r_{sp}E_p\right)\mathbf{\hat{s}}+\left(r_{ps}E_s+r_{pp}E_p\right){\mathbf{\hat{p}}}_r,\\
        \mathbf{E}_t&=\overline{\overline{t}}\mathbf{E}_{i}=\left(t_{ss}E_s+t_{sp}E_p\right)\mathbf{\hat{s}}+\left(t_{ps}E_s+t_{pp}E_p\right){\mathbf{\hat{p}}}_t,
    \end{align}
\end{subequations}
and
\begin{subequations}
    \begin{align}
        \mathbf{H}_i&=\dfrac{1}{\omega\mu_0}\mathbf{k}_i\times\mathbf{E}_{i}=\dfrac{k_1}{\omega\mu_0}\left(E_s\mathbf{\hat{p}}-E_p\mathbf{\hat{s}}\right),\\
        \mathbf{H}_r&=\dfrac{1}{\omega\mu_0}\mathbf{k}_r\times\mathbf{E}_{r}=\dfrac{k_1}{\omega\mu_0}\left[\left(r_{ss}E_s+r_{sp}E_p\right){\mathbf{\hat{p}}}_r-\left(r_{ps}E_s+r_{pp}E_p\right)\mathbf{\hat{s}}\right],\\
        \mathbf{H}_t&=\dfrac{1}{\omega\mu_0}\mathbf{k}_t\times\mathbf{E}_\mathbf{t}=\dfrac{k_2}{\omega\mu_0}\left[\left(t_{ss}E_s+t_{sp}E_p\right){\mathbf{\hat{p}}}_t-\left(t_{ps}E_s+t_{pp}E_p\right)\mathbf{\hat{s}}\right],
    \end{align}
\end{subequations}
where ${k_j}/{\omega\mu_0}=1/\eta_j$ is the characteristic impedance of medium $j$, and
\begin{align}
\overline{\overline{r}}=\left[\begin{matrix}r_{ss}&r_{sp}\\r_{ps}&r_{pp}\\\end{matrix}\right],~~~&
\overline{\overline{t}}=\left[\begin{matrix}t_{ss}&t_{sp}\\t_{ps}&t_{pp}\\\end{matrix}\right],
\end{align}
are the reflection and transmission coefficient tensors (complex quantities).

We next assume that the electromagnetically-thin metasurface residing on the interface ($xy$ plane) between the two semi-infinite media is described by electric and magnetic conductivities $\overline{\overline{\sigma}}$ with in-plane components (no normal current components). The boundary conditions on the metasurface are then
\begin{subequations}\label{eq_s:BC}
    \begin{align}
        \label{eq_s:BC_1}
        \mathbf{\hat{z}}\times\left(\mathbf{E}_1-\mathbf{E}_2\right)&=-\dfrac{1}{2}\overline{\overline{\sigma}}_m\left(\mathbf{H}_1+\mathbf{H}_2\right)_\parallel,\\
        \label{eq_s:BC_2}
        \mathbf{\hat{z}}\times\left(\mathbf{H}_1-\mathbf{H}_2\right)&=\dfrac{1}{2}\overline{\overline{\sigma}}_e\left(\mathbf{E}_1+\mathbf{E}_2\right)_\parallel.
    \end{align}
\end{subequations}
Working on Eq.~\eqref{eq_s:BC_1} we first specify the quantities
\begin{subequations}\label{eq_s:zxE}
    \begin{align}
        \mathbf{\hat{z}}\times\mathbf{E}_i&=-E_s\mathbf{\hat{x}}+E_p\cos{\theta_i}\mathbf{\hat{y}},\\
        \mathbf{\hat{z}}\times\mathbf{E}_r&=-\left(r_{ss}E_s+r_{sp}E_p\right)\mathbf{\hat{x}}-\left(r_{ps}E_s+r_{pp}E_p\right)\cos{\theta_i}\mathbf{\hat{y}},\\
        \mathbf{\hat{z}}\times\mathbf{E}_t&=-\left(t_{ss}E_s+t_{sp}E_p\right)\mathbf{\hat{x}}+\left(t_{ps}E_s+t_{pp}E_p\right)\cos{\theta_t}\mathbf{\hat{y}},
    \end{align}
\end{subequations}
which after some algebra lead to 
\begin{equation}\label{eq_s:sm_1}
    \mathbf{\hat{z}}\times\left(\mathbf{E}_1-\mathbf{E}_2\right)=\mathbf{\hat{z}}\times\left(\mathbf{E}_i+\mathbf{E}_r-\mathbf{E}_t\right)=
    \begin{bmatrix}
        \dfrac{t_{sp}-r_{sp}}{\cos{\theta_i}} & -1-r_{ss}+t_{ss}\\
        1-r_{pp}-t_{pp}\dfrac{\cos{\theta_t}}{\cos{\theta_i}} & -r_{ps}\cos{\theta_i}-t_{ps}\cos{\theta_t}
    \end{bmatrix}
    \begin{bmatrix}
        E_p\cos{\theta_i}\\
        E_s
    \end{bmatrix}.
\end{equation}
Similarly, we find that
\begin{equation}\label{eq_s:sm_2}
    (\mathbf{H}_1+\mathbf{H}_2)_\parallel=
    \begin{bmatrix}
        -\dfrac{1}{\zeta_{1}^{s}}\dfrac{r_{sp}}{\cos{\theta_i}}+\dfrac{1}{\zeta_{2}^{s}}\dfrac{t_{sp}}{\cos{\theta_i}} & \dfrac{1}{\zeta_{1}^{s}}\left(1-r_{ss}\right)+\ \dfrac{1}{\zeta_{2}^{s}}t_{ss} \\ \\
        -\dfrac{1+r_{pp}}{\zeta_{1}^{p}}-\dfrac{1}{\zeta_{2}^{p}}t_{pp}\dfrac{\cos{\theta_t}}{\cos{\theta_i}} & -\dfrac{1}{\zeta_{1}^{p}}r_{ps}\cos{\theta_i}-\dfrac{1}{\zeta_{2}^{p}}t_{ps}\cos{\theta_t}
    \end{bmatrix}
    \begin{bmatrix}
        E_p\cos{\theta_i} \\
        E_s
    \end{bmatrix},
\end{equation}
where $\zeta_{1}^{s}=\eta_1/\cos{\theta_i}$, $\zeta_{2}^{s}=\eta_2/\cos{\theta_t}$, $\zeta_{1}^{p}=\eta_1\cos{\theta_i}$, and $\zeta_{2}^{p}=\eta_2\cos{\theta_t}$ are the wave impedance for the TE ($s$) and TM ($p$) polarization in the respective medium. By substituting Eq.~\eqref{eq_s:sm_1} and \eqref{eq_s:sm_2} back into Eq.~\eqref{eq_s:BC_1}, we arrive at
\begin{multline}
\begin{bmatrix}
        \dfrac{t_{sp}-r_{sp}}{\cos{\theta_i}} & -1-r_{ss}+t_{ss}\\
        1-r_{pp}-\dfrac{\cos{\theta_t}}{\cos{\theta_i}}t_{pp} & -\cos{\theta_i}r_{ps}-\cos{\theta_t}t_{ps}
    \end{bmatrix}
    \begin{bmatrix}
        E_p\cos{\theta_i}\\
        E_s
    \end{bmatrix} = \\ -\dfrac{1}{2}\overline{\overline{\sigma}}_m\begin{bmatrix}
        -\dfrac{1}{\zeta_{1}^{s}}\dfrac{r_{sp}}{\cos{\theta_i}}+\dfrac{1}{\zeta_{2}^{s}}\dfrac{t_{sp}}{\cos{\theta_i}} & \dfrac{1-r_{ss}}{\zeta_{1}^{s}}+\ \dfrac{1}{\zeta_{2}^{s}}t_{ss} \\ \\
        -\dfrac{1+r_{pp}}{\zeta_{1}^{p}}-\dfrac{1}{\zeta_{2}^{p}}\dfrac{\cos{\theta_t}}{\cos{\theta_i}}t_{pp} & -\dfrac{1}{\zeta_{1}^{p}}\cos{\theta_i}r_{ps}-\dfrac{1}{\zeta_{2}^{p}}\cos{\theta_t}t_{ps}
    \end{bmatrix}
    \begin{bmatrix}
        E_p\cos{\theta_i} \\
        E_s
    \end{bmatrix}.
\end{multline}
The above equation has to hold for any incident field, which leads to
\small{
\begin{multline}\label{eq_s:sm_final}
\overline{\overline{\sigma}}_m
    =
    2\begin{bmatrix}
        \dfrac{t_{sp}-r_{sp}}{\cos{\theta_i}} & -1-r_{ss}+t_{ss}\\
        1-r_{pp}-\dfrac{\cos{\theta_t}}{\cos{\theta_i}}t_{pp} & -\cos{\theta_i}r_{ps}-\cos{\theta_t}t_{ps}
    \end{bmatrix}
    \begin{bmatrix}
        \dfrac{1}{\zeta_{1}^{s}\cos{\theta_i}}r_{sp}-\dfrac{1}{\zeta_{2}^{s}\cos{\theta_i}}t_{sp} & -\dfrac{1-r_{ss}}{\zeta_{1}^{s}}-\dfrac{1}{\zeta_{2}^{s}}t_{ss} \\ \\
        \dfrac{1+r_{pp}}{\zeta_{1}^{p}}+\dfrac{1}{\zeta_{2}^{p}}\dfrac{\cos{\theta_t}}{\cos{\theta_i}}t_{pp} & \dfrac{\cos{\theta_i}}{\zeta_{1}^{p}}r_{ps}+\dfrac{\cos{\theta_t}}{\zeta_{2}^{p}}t_{ps}
    \end{bmatrix}^{-1}.
\end{multline}
}
This is Eq.~(5) of the main text that is used to specify the surface magnetic conductivity. 

For the electric conductivity we work in an entirely similar manner. We first calculate the quantities
\begin{subequations}\label{eq_s:zxH}
    \begin{align}
        \mathbf{\hat{z}}\times\mathbf{H}_i&=\dfrac{1}{\omega\mu_0}{\mathbf{\hat{z}}\times\mathbf{k}}_i\times\mathbf{E}_i=\dfrac{1}{\zeta_{1}^{s}}E_s\mathbf{\hat{y}}+\dfrac{1}{\zeta_{1}^{p}}E_p\cos{\theta_i}\mathbf{\hat{x}},\\
        \mathbf{\hat{z}}\times\mathbf{H}_r&=\dfrac{1}{\omega\mu_0}{\mathbf{\hat{z}}\times\mathbf{k}}_r\times\mathbf{E}_r=-\dfrac{1}{\zeta_{1}^{s}}\left(r_{ss}E_s+r_{sp}E_p\right)\mathbf{\hat{y}}+\dfrac{1}{\zeta_{1}^{p}}\left(r_{ps}E_s+r_{pp}E_p\right)\cos{\theta_i}\mathbf{\hat{x}},\\
        \mathbf{\hat{z}}\times\mathbf{H}_t&=\dfrac{1}{\omega\mu_0}{\mathbf{\hat{z}}\times\mathbf{k}}_t\times\mathbf{E}_t=\dfrac{1}{\zeta_{2}^{s}}\left(t_{ss}E_s+t_{sp}E_p\right)\mathbf{\hat{y}}+\dfrac{1}{\zeta_{2}^{p}}\left(t_{ps}E_s+t_{pp}E_p\right)\cos{\theta_t}\mathbf{\hat{x}}.
    \end{align}
\end{subequations}
This leads to 
\begin{equation}\label{eq_s:se_1}
    \mathbf{\hat{z}}\times\left(\mathbf{H}_1-\mathbf{H}_2\right)=
    \begin{bmatrix}
        \dfrac{1+r_{pp}}{\zeta_{1}^{p}}-\dfrac{1}{\zeta_{2}^{p}}\dfrac{\cos{\theta_t}}{\cos{\theta_i}}t_{pp} & \dfrac{\cos{\theta_i}}{\zeta_{1}^{p}}r_{ps}-\dfrac{\cos{\theta_t}}{\zeta_{2}^{p}}t_{ps} \\ 
        -\dfrac{1}{\zeta_{1}^{s}}\dfrac{r_{sp}}{\cos{\theta_i}}-\dfrac{1}{\zeta_{2}^{s}}\dfrac{t_{sp}}{\cos{\theta_i}} & \dfrac{1-r_{ss}}{\zeta_{1}^{s}}-\dfrac{t_{ss}}{\zeta_{2}^{s}}
    \end{bmatrix}
    \begin{bmatrix}
        E_p\cos{\theta_i} \\
        E_s
    \end{bmatrix}.
\end{equation}
We then calculate
\begin{equation}\label{eq_s:se_2}
\left(\mathbf{E}_1+\mathbf{E}_2\right)_\parallel=
    \begin{bmatrix}
        1-r_{pp}+\dfrac{\cos{\theta_t}}{{\cos{\theta}}_i}t_{pp} & -\cos{\theta_i}r_{ps}+\cos{\theta_t}t_{ps} \\ \\
        \dfrac{r_{sp}+t_{sp}}{\cos{\theta_i}} & 1+r_{ss}+t_{ss}
    \end{bmatrix}
    \begin{bmatrix}
        E_p\cos{\theta_i} \\
        E_s
    \end{bmatrix}.
\end{equation}
By substituting Eq.~\eqref{eq_s:se_1} and \eqref{eq_s:se_2} back into Eq.~\eqref{eq_s:BC_2}, we can solve for the electric conductivity and derive the expression (same as Eq.~(4)):
\small{
\begin{equation} \label{eq_s:se_final}
    \overline{\overline{\sigma}}_e = 2 
        \begin{bmatrix}
        \dfrac{1+r_{pp}}{\zeta_{1}^{p}}-\dfrac{1}{\zeta_{2}^{p}}\dfrac{\cos{\theta_t}}{\cos{\theta_i}}t_{pp} & \dfrac{\cos{\theta_i}}{\zeta_{1}^{p}}r_{ps}-\dfrac{\cos{\theta_t}}{\zeta_{2}^{p}}t_{ps} \\ \\
        -\dfrac{1}{\zeta_{1}^{s}\cos{\theta_i}}r_{sp}-\dfrac{1}{\zeta_{2}^{s}\cos{\theta_i}}t_{sp} & \dfrac{1-r_{ss}}{\zeta_{1}^{s}}-\dfrac{t_{ss}}{\zeta_{2}^{s}}
    \end{bmatrix}
        \begin{bmatrix}
        1-r_{pp}+\dfrac{\cos{\theta_t}}{{\cos{\theta}}_i}t_{pp} & -\cos{\theta_i}r_{ps}+\cos{\theta_t}t_{ps} \\ \\
        \dfrac{r_{sp}+t_{sp}}{\cos{\theta_i}} & 1+r_{ss}+t_{ss}
    \end{bmatrix}^{-1}.
\end{equation}
}
Finally, a useful and commonly encountered situation is a metasurface sandwiched between the same dielectrics. For the case of normal incidence Eqs.~\eqref{eq_s:se_final} and \eqref{eq_s:sm_final} can be recast to the simplified expressions:
\small{
\begin{equation}\label{eq_s:sigma_e_simplified}
        \overline{\overline{\sigma}}_e = \dfrac{2}{\eta}\dfrac{        
        \begin{bmatrix}            
        {(1+r_{pp}-t_{pp})(1+r_{ss}+t_{ss})-(r_{ps}-t_{ps})(r_{sp}+t_{sp})} &
         {2(r_{ps}-t_{ps})} \\
           {-2(r_{sp}+t_{sp})} &
         {(1-r_{pp}+t_{pp})(1-r_{ss}-t_{ss})-(r_{ps}-t_{ps})(r_{sp}+t_{sp})}
         \end{bmatrix}}{(1-r_{pp}+t_{pp})(1+r_{ss}+t_{ss})+(r_{ps}-t_{ps})(r_{sp}+t_{sp})},
\end{equation}
}
\small{
\begin{equation}\label{eq_s:sigma_m_simplified}
        \overline{\overline{\sigma}}_m = 2\eta\dfrac{        
        \begin{bmatrix}            
        {(1+r_{pp}+t_{pp})(1+r_{ss}-t_{ss})-(r_{sp}-t_{sp})(r_{ps}+t_{ps})} &
        {-2(r_{sp}-t_{sp})} \\
        {2(r_{ps}+t_{ps})} &
        {(1-r_{pp}-t_{pp})(1-r_{ss}+t_{ss})-(r_{sp}-t_{sp})(r_{ps}+t_{ps})}
         \end{bmatrix}}{(1+r_{pp}+t_{pp})(1-r_{ss}+t_{ss})+(r_{sp}-t_{sp})(r_{ps}+t_{ps})},
\end{equation}
}
where $\eta$ is the characteristic impedance of the uniform host medium.

\subsection{Inverse expressions}
This Section provides a detailed derivation of the reflection $\overline{\overline{r}}$ and transmission $\overline{\overline{t}}$ tensors, given the electric $\overline{\overline{\sigma}}_e$ and magnetic $\overline{\overline{\sigma}}_m$ surface conductivities.
We rewrite Eqs.~\eqref{eq_s:zxE} and \eqref{eq_s:sm_2} into matrix form:
\begin{subequations}\label{eq_s:inv_preliminary_1}
\begin{align}
    \mathbf{\hat{z}}\times\mathbf{E}_i&=\left[\begin{matrix}-1&0\\0&\cos{\theta_i}\\\end{matrix}\right]\left[\begin{matrix}
        E_s\\E_p\end{matrix}\right], \\
    \mathbf{\hat{z}}\times\mathbf{E}_r&=\left[\begin{matrix}-1&0\\0&-\cos{\theta_i}\\\end{matrix}\right]\overline{\overline{r}}\left[\begin{matrix}
        E_s\\E_p\end{matrix}\right], \\
    \mathbf{\hat{z}}\times\mathbf{E}_t&=\left[\begin{matrix}-1&0\\0&\cos{\theta_t}\\\end{matrix}\right]\overline{\overline{t}}\left[\begin{matrix}
        E_s\\E_p\end{matrix}\right],\\
    \left(\mathbf{H}_1+\mathbf{H}_2\right)_\parallel&=\left(\left[\begin{matrix}\dfrac{1}{\zeta_{1}^{s}}&0\\0&-\dfrac{\cos{\theta_i}}{\zeta_{1}^{p}}\\\end{matrix}\right]+\left[\begin{matrix}-\dfrac{1}{\zeta_{1}^{s}}&0\\0&-\dfrac{\cos{\theta_i}}{\zeta_{1}^{p}}\\\end{matrix}\right]\overline{\overline{r}}+\left[\begin{matrix}\dfrac{1}{\zeta_{2}^{s}}&0\\0&-\dfrac{\cos{\theta_t}}{\zeta_{2}^{p}}\\\end{matrix}\right]\overline{\overline{t}}\right)\left[\begin{matrix}
        E_s\\E_p\end{matrix}\right].
\end{align}
\end{subequations}
We note that in the above equations the leftmost matrices in the right-hand-side perform two operations: the vector product and the transformation from the ($\mathbf{\hat{s}}$, $\mathbf{\hat{p}}$) base to the Cartesian system. Next, Eqs.~\eqref{eq_s:inv_preliminary_1} are combined through the first boundary condition Eq.~\eqref{eq_s:BC_1} to give
\begin{multline}
    \label{eq_s:inv_1}
    \left(\left[\begin{matrix}-1&0\\0&-\cos{\theta_i}\\\end{matrix}\right]+\dfrac{1}{2}\overline{\overline{\sigma}}_m\left[\begin{matrix}-\dfrac{1}{\zeta_{1}^{s}}&0\\0&-\dfrac{\cos{\theta_i}}{\zeta_{1}^{p}}\\\end{matrix}\right]\right)\overline{\overline{r}}+\left(\left[\begin{matrix}1&0\\0&-\cos{\theta_t}\\\end{matrix}\right]+\dfrac{1}{2}\overline{\overline{\sigma}}_m\left[\begin{matrix}\dfrac{1}{\zeta_{2}^{s}}&0\\0&-\dfrac{\cos{\theta_t}}{\zeta_{2}^{p}}\\\end{matrix}\right]\right)\overline{\overline{t}} \\ =\left[\begin{matrix}1&0\\0&-\cos{\theta_i}\\\end{matrix}\right]-\dfrac{1}{2}\overline{\overline{\sigma}}_m\left[\begin{matrix}\dfrac{1}{\zeta_{1}^{s}}&0\\0&-\dfrac{\cos{\theta_i}}{\zeta_{1}^{p}}\\\end{matrix}\right].
\end{multline}

We repeat the process for the Eqs.~\eqref{eq_s:zxH} and \eqref{eq_s:se_2}:
\begin{subequations}
    \begin{align}
        \mathbf{\hat{z}}\times\mathbf{H}_i&=\left[\begin{matrix}0&\dfrac{\cos{\theta_i}}{\zeta_{1}^{p}}\\\dfrac{1}{\zeta_{1}^{s}}&0\\\end{matrix}\right]\left[\begin{matrix}
        E_s\\E_p\end{matrix}\right],\\
        \mathbf{\hat{z}}\times\mathbf{H}_r&=\left[\begin{matrix}0&\dfrac{\cos{\theta_i}}{\zeta_{1}^{p}}\\-\dfrac{1}{\zeta_{1}^{s}}&0\\\end{matrix}\right]\overline{\overline{r}}\left[\begin{matrix}
        E_s\\E_p\end{matrix}\right],\\
        \mathbf{\hat{z}}\times\mathbf{H}_t&=\left[\begin{matrix}0&\dfrac{\cos{\theta_t}}{\zeta_{2}^{p}}\\\dfrac{1}{\zeta_{2}^{s}}&0\\\end{matrix}\right]\overline{\overline{t}}\left[\begin{matrix}
        E_s\\E_p\end{matrix}\right],\\        \left(\mathbf{E}_1+\mathbf{E}_2\right)_\parallel&=\left(\left[\begin{matrix}0&\cos{\theta_i}\\1&0\\\end{matrix}\right]+\left[\begin{matrix}0&-\cos{\theta_i}\\1&0\\\end{matrix}\right]\overline{\overline{r}}+\left[\begin{matrix}0&\cos{\theta_t}\\1&0\\\end{matrix}\right]\overline{\overline{t}}\right)\left[\begin{matrix}
        E_s\\E_p\end{matrix}\right].
    \end{align}
\end{subequations}
The above equations are combined through the second boundary condition Eq.~\eqref{eq_s:BC_2} to yield
\begin{multline}\label{eq_s:inv_2}
    \left(\left[\begin{matrix}0&\dfrac{\cos{\theta_i}}{\zeta_{1}^{p}}\\-\dfrac{1}{\zeta_{1}^{s}}&0\\\end{matrix}\right]-\dfrac{1}{2}\overline{\overline{\sigma}}_e\left[\begin{matrix}0&-\cos{\theta_i}\\1&0\\\end{matrix}\right]\right)\overline{\overline{r}}+\left(\left[\begin{matrix}0&-\dfrac{\cos{\theta_t}}{\zeta_{2}^{p}}\\-\dfrac{1}{\zeta_{2}^{s}}&0\\\end{matrix}\right]-\dfrac{1}{2}\overline{\overline{\sigma}}_e\left[\begin{matrix}0&\cos{\theta_t}\\1&0\\\end{matrix}\right]\right)\overline{\overline{t}}\\
    =\left[\begin{matrix}0&-\dfrac{\cos{\theta_i}}{\zeta_{1}^{p}}\\-\dfrac{1}{\zeta_{1}^{s}}&0\\\end{matrix}\right]+\dfrac{1}{2}\overline{\overline{\sigma}}_e\left[\begin{matrix}0&\cos{\theta_i}\\1&0\\\end{matrix}\right].
\end{multline}
Equations \eqref{eq_s:inv_1} and \eqref{eq_s:inv_2} form a linear system of eight equations with eight unknowns and can be solved for $\overline{\overline{r}}$ and $\overline{\overline{t}}$. 

We should note that if either of the conductivities is zero, one can also use these equations to directly find the relationship between the reflection and transmission coefficients. For example, if the electric conductivity is negligible, then Eq.~\eqref{eq_s:inv_2} yields the following
\begin{equation}
    \overline{\overline{r}}+
    \dfrac{\eta_{1}}{\eta_{2}}\left[\begin{matrix}\dfrac{\cos{\theta_t}}{\cos{\theta_i}} & 0\\
    0 & -1\\\end{matrix}\right]\overline{\overline{t}}=
    \left[\begin{matrix}1&0\\0&-1\\\end{matrix}\right].
\end{equation}
Similarly, if the magnetic conductivity is negligible then Eq.~\eqref{eq_s:inv_1} yields
\begin{equation}
    \overline{\overline{r}}+\left[\begin{matrix}-1&0\\0&\dfrac{\cos{\theta_t}}{\cos{\theta_i}}\\\end{matrix}\right]\overline{\overline{t}} =
    \left[\begin{matrix}-1&0\\0&1\\\end{matrix}\right].
\end{equation}

\subsection{Transformation to the Cartesian base}
From Eq.~\eqref{eq_s:unit_vectors} (excluding the $z$ components), Eq.~\eqref{eq_s:r_t_amplitudes} and the fact that $E_s=E_y$, $E_p=E_x/\cos{\theta_i}$ one can directly derive the relationships:
\begin{subequations}\label{eq_s:r_t(xy) -> r_t(sp)}
    \begin{align}
        \left[\begin{matrix}r_{xx}&r_{xy}\\r_{yx}&r_{yy}\\\end{matrix}\right]=\left[\begin{matrix}-r_{pp} & -\cos{\theta_i}r_{ps} \\ \\\dfrac{r_{sp}}{\cos{\theta_i}} & r_{ss} \\ \end{matrix} \right], \\ \notag\\
        \left[\begin{matrix}t_{xx}&t_{xy}\\t_{yx}&t_{yy}\\\end{matrix}\right]=\left[\begin{matrix}\dfrac{\cos{\theta_t}}{\cos{\theta_i}}t_{pp} & \cos{\theta_t}t_{ps} \\ \\ \dfrac{t_{sp}}{\cos{\theta_i}}&t_{ss}\\\end{matrix}\right].
    \end{align}
\end{subequations}
Although straightforward to derive from the above equations, we also denote the useful inverse relations:
\begin{subequations} \label{eq_s:r_t(sp) -> r_t(xy)}
    \begin{align}
        \left[\begin{matrix}r_{ss}&r_{sp}\\r_{ps}&r_{pp}\\\end{matrix}\right]=
        \left[\begin{matrix}r_{yy} & \cos{\theta_i}r_{yx} \\ \\-\dfrac{r_{xy}}{\cos{\theta_i}}&-r_{xx}\\\end{matrix}\right], \\ \notag\\
        \left[\begin{matrix}t_{ss} & t_{sp} \\ t_{ps} & t_{pp} \\ \end{matrix}\right] =
        \left[\begin{matrix}t_{yy} & \cos{\theta_i}t_{yx} \\ \\
        \dfrac{t_{xy}}{\cos{\theta_t}} & \dfrac{\cos{\theta_i}}{\cos{\theta_t}}t_{xx} \\ \end{matrix} \right].
    \end{align}
\end{subequations}
Substituting Eqs.~\eqref{eq_s:r_t(sp) -> r_t(xy)} into Eq.~\eqref{eq_s:sm_final} and \eqref{eq_s:se_final} one finds the following simpler expressions of the electric and magnetic conductivities in terms of the Cartesian reflection and transmission coefficients:
\begin{equation}\label{eq_s:sm_final_xy}
\overline{\overline{\sigma}}_m =
    2\begin{bmatrix}
        -r_{yx}+t_{yx} & -1-r_{yy}+t_{yy}\\
        1+r_{xx}-t_{xx} & r_{xy}-t_{xy}
    \end{bmatrix}
    \begin{bmatrix}
        \dfrac{1}{\zeta_{1}^{s}}r_{yx}-\dfrac{1}{\zeta_{2}^{s}}t_{yx} & -\dfrac{1-r_{yy}}{\zeta_{1}^{s}}-\dfrac{1}{\zeta_{2}^{s}}t_{yy} \\ \\
        \dfrac{1-r_{xx}}{\zeta_{1}^{p}}+\dfrac{1}{\zeta_{2}^{p}}t_{xx} & -\dfrac{1}{\zeta_{1}^{p}}r_{xy}+\dfrac{1}{\zeta_{2}^{p}}t_{xy}
    \end{bmatrix}^{-1},
\end{equation}
\begin{equation} \label{eq_s:se_final_xy}
    \overline{\overline{\sigma}}_e = 2 
        \begin{bmatrix}
        \dfrac{1-r_{xx}}{\zeta_{1}^{p}}-\dfrac{1}{\zeta_{2}^{p}}t_{xx} & -\dfrac{1}{\zeta_{1}^{p}}r_{xy}-\dfrac{1}{\zeta_{2}^{p}}t_{xy} \\ \\
        -\dfrac{1}{\zeta_{1}^{s}}r_{yx}-\dfrac{1}{\zeta_{2}^{s}}t_{yx} & \dfrac{1-r_{yy}}{\zeta_{1}^{s}}-\dfrac{1}{\zeta_{2}^{s}}t_{yy}
    \end{bmatrix}
        \begin{bmatrix}
        1+r_{xx}+t_{xx} & r_{xy}+t_{xy} \\ \\
        r_{yx}+t_{yx} & 1+r_{yy}+t_{yy}
    \end{bmatrix}^{-1}.
\end{equation}

\clearpage
\section{Passivity constraints on the surface conductivity tensors}


A passive system does not generate energy but only dissipates (or stores) it. Assuming time-harmonic fields, the electric power loss density \(P\), averaged over one period, is 
\begin{equation}
\langle P \rangle = \frac{1}{2} \operatorname{Re}\left\{ {\mathbf{E}} \cdot {\mathbf{J}}^* \right\}
= \frac{1}{2} \operatorname{Re}\left\{ {\mathbf{E}}^\dagger {\mathbf{J}} \right\}, 
\end{equation}
where the dagger denotes conjugate transpose. For a system with electric surface conductivity, the surface current density is connected with the tangential electric field through
\begin{equation}
\mathbf{J}_{||} = \overline{\overline{\sigma}}_e \mathbf{E}_{||}.
\end{equation}
An entirely similar discussion can be conducted for the magnetic power density. The time-averaged power density becomes
\begin{equation}
\langle P \rangle
= \frac{1}{2} \operatorname{Re}\left\{ {\mathbf{E}}_{\parallel}^\dagger \,\overline{\overline{\sigma}}_e\, {\mathbf{E}}_{\parallel} \right\}
= \frac{1}{2}
{\mathbf{E}}_{\parallel}^\dagger
\left( \frac{\overline{\overline{\sigma}}_e + \overline{\overline{\sigma}}_e^\dagger}{2} \right)
{\mathbf{E}}_{\parallel},
\end{equation}
i.e., a quadratic form which involves the \textit{Hermitian part} of $\overline{\overline{\sigma}}_e$. The dissipated power should be non-negative for any complex vector ${\mathbf{E}}_{\parallel}$, which is equivalent to requiring that the Hermitian part of $\overline{\overline{\sigma}}_e$ be positive semi-definite \cite{HornJohnson2012}:
\begin{equation}
\frac{\overline{\overline{\sigma}}_e + \overline{\overline{\sigma}}_e^\dagger}{2} \succeq 0.
\end{equation}
For a $2\times 2$ conductivity matrix both eigenvalues should satisfy
\begin{equation}
\lambda_{\pm} \ge 0.
\end{equation}


For a \textit{reciprocal} system, the surface conductivity tensor ($\sigma_{ij} \in \mathbb{C}$) is
\begin{equation}
\overline{\overline{\sigma}}_e = \begin{bmatrix}
\sigma_{xx} & \sigma_{xy} \\
\sigma_{xy} & \sigma_{yy}
\end{bmatrix},
\end{equation}
i.e., $\sigma_{xy} = \sigma_{yx}$ . 
In this case, the Hermitian part of $\overline{\overline{\sigma}}_e$ is given by simply taking the real part of each element: 
\begin{equation}
\begin{bmatrix}
\operatorname{Re}\{\sigma_{xx}\} & \operatorname{Re}\{\sigma_{xy}\} \\
\operatorname{Re}\{\sigma_{xy}\} & \operatorname{Re}\{\sigma_{yy}\}
\end{bmatrix}.
\end{equation}

The eigenvalues of a \(2 \times 2\) matrix  \(M = \begin{bmatrix} a & c \\ c & b \end{bmatrix}\) are:
\begin{equation}
\lambda_{\pm} = \frac{a + b}{2} \pm \sqrt{\left(\frac{a - b}{2}\right)^2 + c^2}
\end{equation}
where \(a = \operatorname{Re}\{\sigma_{xx}\}\), \(b = \operatorname{Re}\{\sigma_{yy}\}\), and \(c = \operatorname{Re}\{\sigma_{xy}\}\). For the matrix to be positive semi-definite, we require the smaller eigenvalue \(\lambda_{-}\) to be greater than or equal to zero
\begin{equation}
\lambda_{-} = \frac{a + b}{2} - \sqrt{\left(\frac{a - b}{2}\right)^2 + c^2} \geq 0.
\end{equation}
After some algebra we reach 
\begin{equation}
ab \geq c^2.
\end{equation}

In summary, we need the following conditions to hold for the surface conductivity tensor: 

\begin{itemize}
    \item The diagonal elements must be non-negative:
\end{itemize}
    \begin{equation}
    \operatorname{Re}\{\sigma_{xx}\} \geq 0, \quad \operatorname{Re}\{\sigma_{yy}\} \geq 0
    \end{equation}

\begin{itemize}
\item The real part of the off-diagonal element squared should be smaller than the product of the diagonal elements:
\end{itemize}
    \begin{equation} 
    \operatorname{Re}\{\sigma_{xy}\}^2 \leq \operatorname{Re}\{\sigma_{xx}\} \operatorname{Re}\{\sigma_{yy}\}  
    \end{equation}
This result agrees with Ref.~\onlinecite{Arnold2025} starting from a different premise. 

\clearpage
\section{Output polarization assessment. Example of cut-wire pair for $y$-polarized incidence.}

\begin{figure}[h]
    \centering
    \includegraphics[alt={Polarization conversion for a plane wave transmitted through the cut-wire-pair meta-atom at forty-five-degree rotation under normal incidence and assuming linear y-polarized illumination. The transmitted polarization state is mapped on the Poincare sphere and also described via the axial ratio and phase difference spectra for the two orthogonal linear components.}]{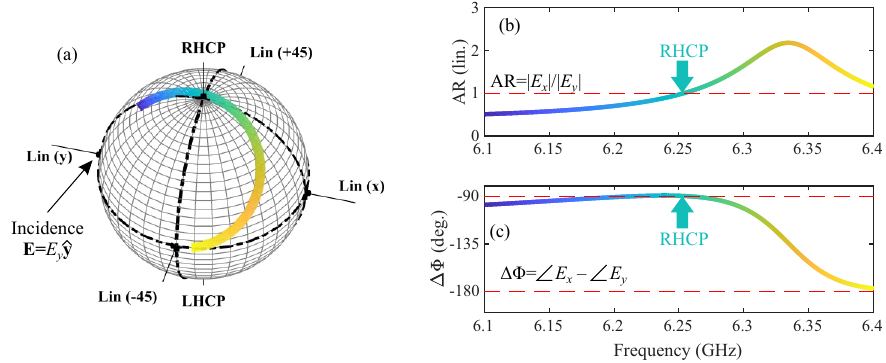}
    \caption{{Polarization conversion for a plane wave transmitted through the cut-wire-pair meta-atom (rotation $\psi=\pi/4$), when $\theta=0$ and assuming linear $y$-polarized illumination. (a) Transmitted polarization state mapped on the Poincar\'{e} sphere for the frequency range 6.1-6.4~GHz. (b) Axial ratio and (c) phase difference spectra for the transmitted plane wave. The transmitted polarization is almost perfectly RHCP at 6.25~GHz and it approaches the linear $-45^\circ$ state at 6.4~GHz.}}
    \label{fig:polarization_conv} 
\end{figure}

Anisotropic metasurfaces can be used for polarization conversion, e.g., linear-to-circular (flat equivalent of a quarter-wave plate). This has been exemplified with the cut-wire pair meta-atom in Section III.B of the main paper. Specifically, Fig.~8 studies the output polarization for $x$-polarized incidence. In Fig.~\ref{fig:polarization_conv}, we show a similar plot for $y$-polarized incidence. It can be seen that incident $y$  polarization is converted to LHCP at the frequency of 6.25~GHz (instead of RHCP). The polarization state is clearly shown on the Poincar\'{e} sphere [Fig.~\ref{fig:polarization_conv}(a)]. It can be also described by two scalar values, the axial ratio $\text{AR}=|E_x|/|E_y|$ [Fig.~\ref{fig:polarization_conv}(b)] and the phase difference $\Delta\Phi=\angle{E_x}-\angle{E_y}\in[-\pi,+\pi]$, Fig.~\ref{fig:polarization_conv}(c). Note that the handedness of the elliptical polarizations is defined by $\Delta\Phi>0$ (left) or $<0$ (right), respectively. 


\clearpage
\section{\label{Section:Lin2dCirc}Polarization state conversion: Dual linear to circular}

Using the proposed formulation, we will study the conversion of dual linear polarization to circular (i.e., simultaneously $s$ to LHCP and $p$ to RHCP, or vice versa) and specifically the required form of the magnetic/electric conductivity for conversion to take place. To the left and right of the metasurface the medium is isotropic with characteristic impedance $\eta$ and we assume normal incidence. The polarization conversion can be achieved in either reflection or transmission. The necessary and sufficient conditions for this are
\begin{subequations}
\label{eq:pol_conversion_conditions}
\begin{align}
    r_{ps}=ir_{ss}, r_{pp}=-ir_{sp}, \\
    t_{ps}=it_{ss}, t_{pp}=-it_{sp},
\end{align}
\end{subequations}
for reflection and transmission, respectively.

For a purely magnetic meta-atom, we solve Eqs.~\eqref{eq_s:inv_1} and \eqref{eq_s:inv_2} for the reflection and transmission coefficients and then substitute the electric conductivity components with zeros. Then, for operation in reflection or transmission we substitute the corresponding conditions from Eq.~\eqref{eq:pol_conversion_conditions}. In reflection the magnetic conductivities take the form 
\begin{subequations}
    \begin{align}
        \frac{\sigma_m^{xx}}{2\eta_1}&=-\frac{\eta_2-\eta_1 +2(\eta_1+\eta_2)r_{ps}r_{sp}+(\eta_1+\eta_2){i}r_{ps}-(\eta_2-\eta_1){i}r_{sp}}{2\eta_1(2r_{ps}r_{sp}+1+{i}r_{ps}-{i}r_{sp})},\\
        \frac{\sigma_m^{xy}}{2\eta_1}&=-\frac{r_{sp}}{2r_{ps}r_{sp}+1+{i}r_{ps}-{i}r_{sp}},\\
        \frac{\sigma_m^{yx}}{2\eta_1}&=\frac{r_{ps}}{2r_{ps}r_{sp}+1+{i}r_{ps}-{i}r_{sp}},\\
        \frac{\sigma_m^{yy}}{2\eta_1}&=-\frac{\eta_2-\eta_1+2(\eta_1+\eta_2)r_{ps}r_{sp}+(\eta_2-\eta_1){i}r_{ps}-(\eta_1+\eta_2){i}r_{sp}}{2\eta_1(2r_{ps}r_{sp}+1+{i}r_{ps}-{i}r_{sp})},
    \end{align}
\end{subequations}
while for transmission they are found to be
\begin{subequations}
    \begin{align}
        \frac{\sigma_m^{xx}}{2\eta_2}&=-\frac{(\eta_1+\eta_2)t_{ps}-\eta_2 {i}}{{2\eta_2}t_{ps}},\\ 
        \frac{\sigma_m^{xy}}{2\eta_2}&=-\frac{1 }{2t_{ps}},\\ 
        \frac{\sigma_m^{yx}}{2\eta_2}&=-\frac{1 }{2t_{sp}},\\ 
        \frac{\sigma_m^{yy}}{2\eta_2}&=-\frac{(\eta_1+\eta_2)t_{sp}-\eta_2 {i}}{{2\eta_2}t_{sp}}.
    \end{align}
\end{subequations}
Using Eqs.~\eqref{eq_s:inv_1} and \eqref{eq_s:inv_2}, we can also find the relation connecting the conductivity components. For the case of a symmetric meta-atom where $\sigma_m^{xx}=\sigma_m^{yy}$ and operation in reflection, this relationship takes the form
\begin{subequations}
\begin{align}
    \sigma_m^{xy}&=\sigma_m^{yx},\\
    \label{eq_s:magnetic conductivity reflection criterion}
    \left(\frac{\sigma_m^{yx}}{2\eta_2}-\frac{\eta_1}{2\eta_2}{i}\right)^2-\frac{-2{\eta_1 }^2 +{\eta_2 }^2}{4\eta_2^2} -\frac{\sigma_m^{xx}}{2\eta_2} -\left(\frac{\sigma_m^{xx}}{2\eta_2}\right)^2&=0,
\end{align}
\end{subequations}
while in transmission the relation becomes
\begin{subequations}
\begin{align}
    \sigma_m^{xy}&=\sigma_m^{yx},\\
    \label{eq_s:magnetic conductivity transmission criterion}
    \frac{\sigma_m^{yx}}{2\eta_2}-\frac{\eta_1 +\eta_2}{2\eta_2}i-\frac{\sigma_m^{xx}}{2\eta_2}i&=0.
\end{align}
\end{subequations}

For a purely electric meta-atom we follow the same procedure but set the magnetic conductivity to zero, which leads to the following expressions for the electric conductivity for operation in reflection
\begin{subequations}
    \begin{align}
        \frac{\sigma_e^{xx}\eta_1}{2}&=-\frac{-(\eta_2 -\eta_1) +2(\eta_1+\eta_2) r_{ps}r_{sp}+(\eta_2-\eta_1)ir_{ps}+(\eta_1 +\eta_2)ir_{sp}}{2\eta_2 {\left(2r_{ps}r_{sp}+1-r_{ps}{i}+r_{sp}{i}\right)}},\\
        \frac{\sigma_e^{xy}\eta_1}{2}&=\frac{r_{ps}}{2r_{ps}r_{sp}+1-r_{ps}{i}+r_{sp}{i}},\\
       \frac{\sigma_e^{yx}\eta_1}{2}&=-\frac{r_{sp}}{2r_{ps}r_{sp}+1-r_{ps}{i}+r_{sp}{i}},\\
        \frac{\sigma_e^{yy}\eta_1}{2}&=-\frac{-(\eta_2-\eta_1)+2(\eta_1+\eta_2)r_{ps}r_{sp}-(\eta_1+\eta_2)ir_{ps}-(\eta_2-\eta_1)ir_{sp}}{2\eta_2 {\left(2r_{ps}r_{sp}+1-r_{ps}{i}+r_{sp}{i}\right)}}.
    \end{align}
\end{subequations}
For operation in transmission the electric conductivity is found to be 
\begin{subequations}
    \begin{align}
        \frac{\sigma_e^{xx}\eta_1}{2}&=-\frac{(\eta_1+\eta_2)t_{sp}-\eta_2 {i}}{2 \eta_2 t_{sp}},\\
        \frac{\sigma_e^{xy}\eta_1}{2}&=\frac{1}{2 t_{sp}},\\
        \frac{\sigma_e^{yx}\eta_1}{2}&=\frac{1}{2 t_{ps}},\\
        \frac{\sigma_e^{yy}\eta_1}{2}&=-\frac{(\eta_1+\eta_2)t_{ps}-\eta_2 {i}}{2\eta_2 t_{ps}}.
    \end{align}
\end{subequations}
Again, using the inverse relationships for a symmetric meta-atom $\sigma_e^{xx}=\sigma_e^{yy}$ we arrive at the following relations between conductivity components: For operation in reflection
\begin{subequations}
\begin{align}
\sigma_e^{xy}&=\sigma_e^{yx},\\
\label{eq_s:electric conductivity reflection criterion}
\left(\frac{\sigma_e^{yx}\eta_1}{2}+\frac{1}{2}i\right)^2-\left( \frac{\sigma_e^{xx}\eta_1}{2} \right)^2 -\frac{\eta_1}{\eta_2} \frac{\sigma_e^{xx}\eta_1}{2} -\frac{{\eta_1 }^2 -2{\eta_2 }^2}{4\eta_2^2}&=0,
\end{align}
\end{subequations}

and for operation in transmission
\begin{subequations}
\begin{align}
    \sigma_e^{xy}&=\sigma_e^{yx},\\
    \label{eq_s:electric conductivity transmission criterion}
    \frac{\sigma_e^{yx}\eta_1}{2}+\frac{{\eta_1 +\eta_2}}{2\eta_2 }i +\frac{\sigma_e^{xx}\eta_1}{2}i&=0.
\end{align}
\end{subequations}

\begin{figure}[t]
    \centering
    \includegraphics[width=0.85\textwidth,alt={Calculation of criterion for dual linear to circular polarization conversion in the tilted wire-pair meta-atom and the tilted wire meta-atom. The real and imaginary parts do not zero-out simultaneously for the same frequency and, thus, the criteria are not satisfied in any of the two cases.}]{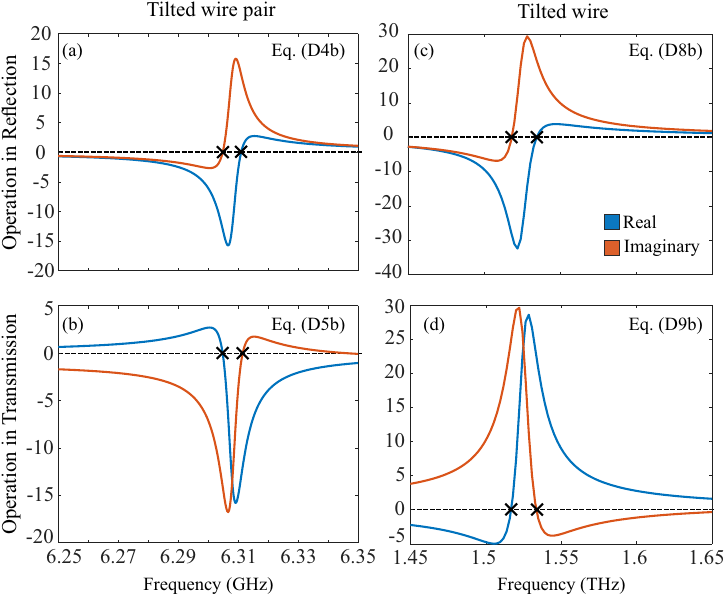}
    \caption{(a),(b) Calculation of Eq.~\eqref{eq_s:magnetic conductivity reflection criterion} and Eq.~\eqref{eq_s:magnetic conductivity transmission criterion} for the tilted wire-pair meta-atom of the main text. We can see  the real and imaginary parts do not zero-out simultaneously for the same frequency. The additional effect of a non-zero electric conductivity finally enables the linear to circular polarization conversion in transmission, as shown in Fig.~\ref{fig:polarization_conv}. (c),(d) Calculation of Eq.~\eqref{eq_s:electric conductivity reflection criterion} criterion and \eqref{eq_s:electric conductivity transmission criterion} for the tilted wire meta-atom of the main text. The criteria are again never satisfied, since no polarization conversion was observed.}
    \label{fig_s:polarization_condition}
\end{figure}

Finally, we applied Eqs.~\eqref{eq_s:magnetic conductivity reflection criterion}, \eqref{eq_s:magnetic conductivity transmission criterion} and Eqs.~\eqref{eq_s:electric conductivity reflection criterion}, \eqref{eq_s:electric conductivity transmission criterion} to the retrieved conductivity of the magnetic and electric meta-atoms, respectively. We first check Eq.~\eqref{eq_s:magnetic conductivity reflection criterion} and Eq.~\eqref{eq_s:magnetic conductivity transmission criterion} for the tilted wire pair in reflection and transmission, respectively. We already know from Fig.~\ref{fig:polarization_conv} that the tilted wire pair example can convert linear to circular polarization in transmission. Fig.~\ref{fig_s:polarization_condition}(a),(b) shows the real and imaginary parts of Eq.~\eqref{eq_s:magnetic conductivity reflection criterion} and Eq~\eqref{eq_s:magnetic conductivity transmission criterion}. These criteria are satisfied when the blue and red curves simultaneously cross the zero line, which in this example never strictly happens. However, the tilted wire pair has an electric resonance whose tail produces a non-negligible electric conductivity in the vicinity of the magnetic resonance. With the additional effect of the non-zero electric conductivity, we were able to achieve dual linear to circular polarization conversion at 6.25~GHz (see Fig. 8). 

We now look into the tilted wire example which, as opposed to the case of the tilted-wire pair (not purely magnetic), is a purely electric meta-atom. Thus, if there is polarization conversion this should be reflected in our criteria. In Fig.~\ref{fig_s:polarization_condition}(c),(d) we plot the real and imaginary part of Eq.~\eqref{eq_s:electric conductivity reflection criterion} and Eq.~\eqref{eq_s:electric conductivity transmission criterion}. The real and imaginary parts are not simultaneously satisfied in this frequency range. This agrees with the fact that we do not observe linear to circular polarization conversion taking place.


\clearpage
\section{\label{Section:Lin2Circ}Polarization State Conversion: Single Linear to Circular}

\begin{figure}
    \centering
    \includegraphics[width=0.85\textwidth, alt={Calculation of criterion for linear to circular polarization conversion in the case of the tilted wire meta-atom. The criteria are not satisfied at the studied frequency range, which agrees with the fact that there was no observed linear to circular polarization conversion.}]{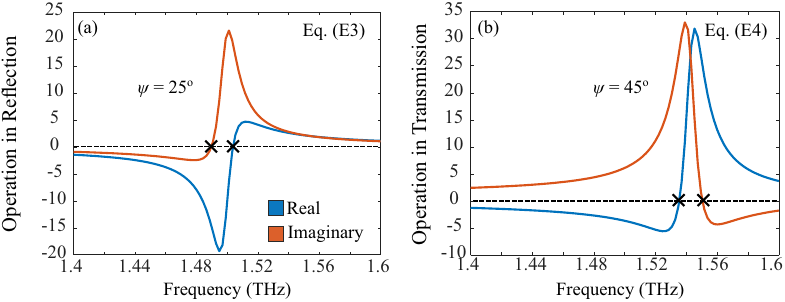}
    \caption{(a),(b) Calculation of Eq.~\eqref{eq_s:purely_elec_crit_no_dual_reflection} and Eq.~\eqref{eq_s:purely_elec_crit_no_dual_transmission}, respectively, for the tilted wire meta-atom of the main text. The conductivities used correspond to normal incidence and a tilt rotation of the atom $\psi=25^\circ$ for operation in reflection and $\psi=45^\circ$ for operation in transmission, respectively. These tilt angles were chosen by inspecting the scattering parameters and selecting the point that resulted into output polarization closest to circular. The criteria are not satisfied at the studied frequency range, which agrees with the fact that there was no observed linear to circular polarization conversion.}
    \label{fig_s:pol_conv_criteria_no_dual}
\end{figure}

In this Section, we repeat the steps of Section~\ref{Section:Lin2dCirc} but without the requirement that both linear polarizations are simultaneously converted to circular. In the following we focus on an incident $s$ polarization. First we look into a purely electric meta atom. For operation in reflection it is required that $ r_{ps}=ir_{ss}$ and the conductivities are given by
\begin{subequations}
    \begin{align}
        \dfrac{\sigma_e^{xx}\eta_1}{2}&=-\dfrac{(\eta_2 -\eta_1)(-1+ir_{ps})+(\eta_1+\eta_2)(r_{ps}r_{sp}-r_{pp}+ir_{pp}r_{ps})}{2\eta_2  {\left(r_{ps} r_{sp}-r_{pp}+1-ir_{ps}+ir_{pp} r_{ps}\right)}},\\
        \dfrac{\sigma_e^{xy}\eta_1}{2}&=\dfrac{r_{ps}}{r_{ps} r_{sp}-r_{pp}+1-ir_{ps}+ir_{pp} r_{ps}},\\
        \dfrac{\sigma_e^{yx}\eta_1}{2}&=-\dfrac{r_{sp}}{r_{ps} r_{sp}-r_{pp}+1-ir_{ps}+ir_{pp} r_{ps}},\\
        \dfrac{\sigma_e^{yy}\eta_1}{2}&=-\dfrac{(\eta_2-\eta_1)(-1+r_{pp})+(\eta_1+\eta_2)(r_{ps}r_{sp}-ir_{ps}+ir_{pp}r_{ps})}{2\eta_2{\left(r_{ps} r_{sp}-r_{pp}+1-ir_{ps}+ir_{pp} r_{ps}\right)}}.
    \end{align}
\end{subequations}
For operation in transmission the requirement is that $t_{ps}=it_{ss}$ and the conductivity components are given by
\begin{subequations}
    \begin{align}
        \dfrac{\sigma_e^{xx}\eta_1}{2}&=-\frac{-2i\eta_2+(\eta_1+\eta_2)(t_{sp}+it_{pp})}{2 \eta_2 {\left(t_{sp}+it_{pp}\right)}},\\
        \dfrac{\sigma_e^{xy}\eta_1}{2}&=\frac{1}{t_{sp}+it_{pp}},\\
        \dfrac{\sigma_e^{yx}\eta_1}{2}&=\frac{t_{sp}}{t_{ps}{\left(t_{sp}+it_{pp}\right)}},\\
        \dfrac{\sigma_e^{yy}\eta_1}{2}&=-\frac{2\eta_2 t_{pp}+(\eta_1+\eta_2)t_{ps}(t_{sp}+it_{pp})}{2\eta_2 t_{ps}{\left(t_{sp}+it_{pp}\right)}}.
    \end{align}
\end{subequations}
Note that for a reciprocal device $\sigma_e^{yx}=\sigma_e^{yx}$ which leads to $r_{ps}=-r_{sp}$ in reflection and $t_{sp}=t_{ps}$ in transmission. The above equations can also be formulated into necessary conductivity criteria for polarization conversion. For operation in reflection we find
\begin{equation}
    \label{eq_s:purely_elec_crit_no_dual_reflection}
    \frac{\sigma_e^{xy}\eta_1}{2} {\left(\frac{\sigma_e^{yx}\eta_1}{2} +i\right)}{\eta_2 }^2-{\left(\frac{\eta_1 -\eta_2}{2} +\eta_2\frac{\sigma_e^{yy}\eta_1}{2} \right)}{\left(\frac{\eta_1 +\eta_2}{2} +\eta_2\frac{\sigma_e^{xx}\eta_1}{2} \right)}=0
\end{equation}
and for operation in transmission we have
\begin{equation}
    \label{eq_s:purely_elec_crit_no_dual_transmission}
    \frac{\sigma_e^{xy}\eta_1}{2} +i(\frac{\eta_1+\eta_2}{2\eta_2}+\frac{\sigma_e^{xx}\eta_1}{2})=0.
\end{equation}
We now apply the above necessary criteria of Eq.~\eqref{eq_s:purely_elec_crit_no_dual_reflection} and \eqref{eq_s:purely_elec_crit_no_dual_transmission} for the purely electric meta-atom of the main text, the tilted wire. The results are shown in Fig.~\ref{fig_s:pol_conv_criteria_no_dual} for a tilt angle of $25^\circ$ and $45^\circ$ for operation in reflection and transmission, respectively. The specific tilted wire cannot convert incident $s$ polarization to circular, which is verified by looking at the scattering parameters (not shown). 

We next look into a purely magnetic meta atom. The requirements on reflection and transmission are the same as in the previous case. For operation in reflection the conductivities are found to be
\begin{subequations}
    \begin{align}
        \dfrac{\sigma_m^{xx}}{2\eta_1}&=-\frac{(\eta_2 -\eta_1)(1-r_{pp})+(\eta_1+\eta_2)(r_{ps}r_{sp}+ir_{ps}+ir_{pp}r_{ps})}{2\eta_1\left(r_{pp}+r_{ps}r_{sp}+1+ir_{ps}+ir_{pp}r_{ps}\right)},\\
        \dfrac{\sigma_m^{xy}}{2\eta_1}&=-\frac{r_{sp}}{r_{pp}+r_{ps}r_{sp}+1+ir_{ps}+ir_{pp}r_{ps}},\\
        \dfrac{\sigma_m^{yx}}{2\eta_1}&=\frac{r_{ps}}{r_{pp}+r_{ps}r_{sp}+1+ir_{ps}+ir_{pp}r_{ps}},\\
        \dfrac{\sigma_m^{yy}}{2\eta_1}&=-\frac{(\eta_2 -\eta_1)(1+ir_{ps}) +(\eta_1+\eta_2)(r_{pp}+r_{ps}r_{sp}+ir_{pp}r_{ps})}{2\eta_1\left(r_{pp}+r_{ps}r_{sp}+1+ir_{ps}+ir_{pp}r_{ps}\right)},
    \end{align}
\end{subequations}
while for operation in transmission the conductivity components are 
\begin{subequations}
    \begin{align}
        \dfrac{\sigma_m^{xx}}{2\eta_2}&=-\frac{2\eta_2 t_{pp}+(\eta_1+\eta_2)t_{ps}(t_{sp}+it_{pp})}{2\eta_2 t_{ps}{\left(t_{sp}+it_{pp}\right)}},\\
        \dfrac{\sigma_m^{xy}}{2\eta_2}&=-\frac{t_{sp}}{t_{ps}{\left(t_{sp}+it_{pp}\right)}},\\
        \dfrac{\sigma_m^{yx}}{2\eta_2}&=-\frac{1}{t_{sp}+it_{pp}},\\
        \dfrac{\sigma_m^{yy}}{2\eta_2}&=-\frac{-i2\eta_2+(\eta_1+\eta_2)(t_{sp}+it_{pp})}{2\eta_2\left(t_{sp}+it_{pp}\right)},
    \end{align}
\end{subequations}
As in the previous case, in a reciprocal device $r_{ps}=-r_{sp}$ in reflection and $t_{sp}=t_{ps}$ in transmission. The necessary criteria for polarization conversion for the purely magnetic meta atom are for operation in reflection
\begin{equation}
    \frac{\sigma_m^{xy}}{2\eta_1}\left(-\frac{\sigma_m^{xy}}{2\eta_1} +i\right)+i{\left(\frac{\eta_1+\eta_2}{2\eta_1} +\frac{\sigma_m^{yy}}{2\eta_1} \right)}{\left(\frac{\eta_2-\eta_1}{2\eta_1}+\frac{\sigma_m^{xx}}{2\eta_1} \right)}=0
\end{equation}
and for operation in transmission
\begin{equation}
    \frac{\sigma_m^{xy}}{2\eta_1}-i\left(\frac{\eta_1+\eta_2}{2\eta_1}+\frac{\sigma_m^{yy}}{2\eta_1}\right)=0.
\end{equation}

\clearpage
\section{Retrieved Electric Conductivity for Cut-Wire-Pair Meta-Atom}
\begin{figure}[h]
    \centering
    \includegraphics[alt={Retrieved electric surface conductivity components of the cut-wire-pair meta-atom under normal incidence for different meta-atom rotations. A broad frequency range is covered. The frequency of the narrowband magnetic surface conductivity resonance is marked with a blue arrow.}]{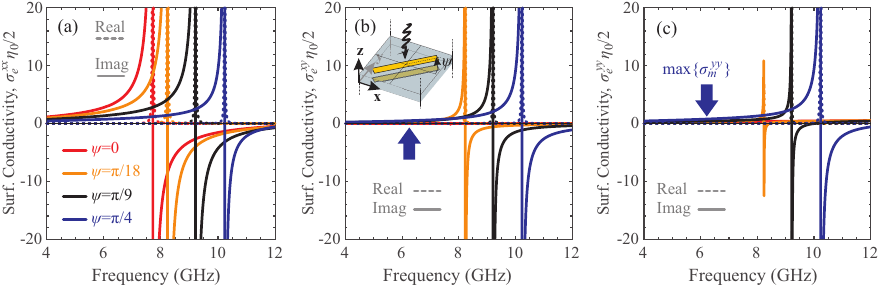}
    \caption{Retrieved normalized (dimensionless) electric surface conductivity components of the cut-wire-pair meta atom at $(\theta_i,\varphi_i)=(0,0)$ illumination for different rotations $\psi$ in a broad band.(a)~$\sigma_e^{xx}$, (b) $\sigma_e^{xy}=\sigma_e^{yx}$, and (c)~$\sigma_e^{yy}$. The frequency of the narrowband magnetic surface conductivity resonance for $\psi=\pi/4$, as shown in {Fig.~6} of the main text, is marked with a blue arrow.}
    \label{fig:SigmeElec_of_CutWirePair}
\end{figure}

The cut-wire-pair meta-atom, apart from a resonance of its retrieved magnetic surface conductivity that arises due to excited counter-propagating currents on the two wires, can also support electric surface conductivity resonances. In Fig.~\ref{fig:SigmeElec_of_CutWirePair} we have retrieved the components of the anisotropic $\overline{\overline{\sigma}}_e$ in a broad band, from 4 to 12~GHz. Evidently, this meta-atom also supports a bright electric resonance (co-propagating currents in the wire-pair), at frequencies near 8-10~GHz.

\end{document}